%% file: tsallis.tex
\documentclass[11pt,a4paper]{article}

\input{no-revtex}

\input{jm-tex-macros-public}
\usepackage[utf8]{inputenc}
\usepackage{comment}
\usepackage{float}
\usepackage{xcolor}
\usepackage{physics}
\usepackage{array}
\usepackage{ragged2e}
\usepackage{booktabs}
\usepackage{latexsym}
\usepackage{mathtools}

\usepackage{subcaption}
\usepackage{graphicx}
\usepackage{subcaption}
\usepackage{cleveref}
\usepackage{amsfonts,amsmath,bm}
\usepackage[toc]{appendix}
\usepackage{subfloat}
\usepackage{indentfirst}

\numberwithin{equation}{section}
\definecolor{darkgreen}{rgb}{0,0.5,0}
\definecolor{darkblue}{rgb}{0,0,0.6}

\hypersetup{
	pdftitle={A q-Inspired Radial Quantum Kinematics},
	pdfauthor={Mir Afrasiar and XianHui Ge},
	pdfsubject={q-inspired radial quantum kinematics},
	pdfkeywords={q-calculus, radial kinematics, self-adjoint momentum, Robertson uncertainty, phase-space measure},
	citecolor=blue,
	colorlinks=true,
	filecolor=red,
	linkcolor=blue,
	linktocpage=true,
	urlcolor=darkgreen
}

\newcount\hour
\newcount\minute
\newtoks\amorpm
\hour=\time\divide\hour by60
\minute=\time{\multiply\hour by60 \global\advance\minute by-\hour}
\edef\standardtime{{\ifnum\hour<12 \global\amorpm={am}%
		\else\global\amorpm={pm}\advance\hour by-12 \fi
		\ifnum\hour=0 \hour=12 \fi
		\number\hour:\ifnum\minute<10 0\fi\number\minute\the\amorpm}}
\edef\militarytime{\number\hour:\ifnum\minute<10 0\fi\number\minute}

\def\draftlabel#1{{\@bsphack\if@filesw {\let\thepage\relax
			\xdef\@gtempa{\write\@auxout{\string
					\newlabel{#1}{{\@currentlabel}{\thepage}}}}}\@gtempa
		\if@nobreak \ifvmode\nobreak\fi\fi\fi\@esphack}
	\gdef\@eqnlabel{#1}}
\def\@eqnlabel{}
\def\@vacuum{}
\def\draftmarginnote#1{\marginpar{\raggedright\scriptsize\tt#1}}

\def\draft{\oddsidemargin -.5truein
		\def\@oddfoot{\sl preliminary draft \hfil
			\rm\thepage\hfil\sl\today\quad\militarytime}
		\let\@evenfoot\@oddfoot \overfullrule 3pt
		\let\label=\draftlabel
		\let\marginnote=\draftmarginnote
		\def\@eqnnum{(\theequation)\rlap{\kern\marginparsep\tt\@eqnlabel}%
			\global\let\@eqnlabel\@vacuum}  }
	
	\def\preprint{\twocolumn\sloppy\flushbottom\parindent 2em
		\leftmargini 2em\leftmarginv .5em\leftmarginvi .5em
		\oddsidemargin -.5in    \evensidemargin -.5in
		\columnsep .4in \footheight 0pt
		\textwidth 10.in        \topmargin  -.4in
		\headheight 12pt \topskip .4in
		\textheight 6.9in \footskip 0pt
		\def\@oddhead{\thepage\hfil\addtocounter{page}{1}\thepage}
		\let\@evenhead\@oddhead \def\@oddfoot{} \def\@evenfoot{} }
	
	\def\numberbysection{\@addtoreset{equation}{section}
		\def\theequation{\thesection.\arabic{equation}}}
	
	\def\underline#1{\relax\ifmmode\@@underline#1\else
		$\@@underline{\hbox{#1}}$\relax\fi}

	\def\titlepage{\@restonecolfalse\if@twocolumn\@restonecoltrue\onecolumn
		\else \newpage \fi \thispagestyle{empty}\c@page\z@
		\def\thefootnote{\fnsymbol{footnote}} }
	
	\def\endtitlepage{\if@restonecol\twocolumn \else \newpage \fi
		\def\thefootnote{\arabic{footnote}}
		\setcounter{footnote}{0}}  

\catcode`@=12
\relax

\def\figcap{\section*{Figure Captions\markboth
		{FIGURECAPTIONS}{FIGURECAPTIONS}}\list
	{Figure \arabic{enumi}:\hfill}{\settowidth\labelwidth{Figure
			999:}
		\leftmargin\labelwidth
		\advance\leftmargin\labelsep\usecounter{enumi}}}
 \relax
\def\tablecap{\section*{Table Captions\markboth
		{TABLECAPTIONS}{TABLECAPTIONS}}\list
	{Table \arabic{enumi}:\hfill}{\settowidth\labelwidth{Table
			999:}
		\leftmargin\labelwidth
		\advance\leftmargin\labelsep\usecounter{enumi}}}
 \relax
\def\reflist{\section*{References\markboth
		{REFLIST}{REFLIST}}\list
	{[\arabic{enumi}]\hfill}{\settowidth\labelwidth{[999]}
		\leftmargin\labelwidth
		\advance\leftmargin\labelsep\usecounter{enumi}}}
 \relax
\makeatletter
\newcounter{pubctr}
\def\publist{\@ifnextchar[{\@publist}{\@@publist}}
\def\@publist[#1]{\list
	{[\arabic{pubctr}]\hfill}{\settowidth\labelwidth{[999]}
		\leftmargin\labelwidth
		\advance\leftmargin\labelsep
		\@nmbrlisttrue\def\@listctr{pubctr}
		\setcounter{pubctr}{#1}\addtocounter{pubctr}{-1}}}
\def\@@publist{\list
	{[\arabic{pubctr}]\hfill}{\settowidth\labelwidth{[999]}
		\leftmargin\labelwidth
		\advance\leftmargin\labelsep
		\@nmbrlisttrue\def\@listctr{pubctr}}}
 \relax
\makeatother
\newskip\humongous \humongous=0pt plus 1000pt minus 1000pt

\newif\ifdtup

\relax

\def\be{\begin{equation}}
	\def\ee{\end{equation}}
\def\ba{\begin{eqnarray}}
	\def\ea{\end{eqnarray}}

\def\IR{\relax{\rm I\kern-.18em R}}

\def\IR{\relax{\rm I\kern-.18em R}}
\def\IL{\relax{\rm I\kern-.18em L}}

\def\inv{^{\raise.15ex\hbox{${\scriptscriptstyle -}$}\kern-.05em 1}}

\def\bea{\begin{eqnarray}}
	\def\eea{\end{eqnarray}}

\definecolor{markcolor2}{rgb}{1,0,0}

\definecolor{markcolor3}{rgb}{0,1,0}

\begin{document}
	
	\thispagestyle{empty}

	\title{\Large\bf\boldmath A $q$-Inspired Radial Quantum Kinematics}

\vspace{0pt}	

\author{
Mir Afrasiar${}^{1}$ and
Xian-Hui Ge${}^{1}$

}
\address{
{\it ${}^1$Department of Physics, Shanghai University, 99 Shangda Road, Shanghai, 200444, China\\}
}

\date{}			

\email{\href{mailto:mirafrasiar@shu.edu.cn}{mirafrasiar@shu.edu.cn}, \href{mailto:gexh@shu.edu.cn}{gexh@shu.edu.cn}
}

\abstract{
We study how the one-dimensional deformation suggested by nonadditive $q$-calculus can be extended to a rotationally covariant quantum model and what observable consequences follow from this extension. We choose a radial extension about a distinguished center in which only the radial derivative is deformed, while tangential derivatives retain their ordinary form. Within this model, we construct self-adjoint frame-momentum operators and derive a rotationally invariant uncertainty bound with a correction proportional to the mean distance from the center. We then adopt a non-negative kinetic Hamiltonian built from these momenta and derive the leading harmonic-oscillator energy shifts for weak deformation. The energy corrections split the levels with different orbital angular momenta that are degenerate in the undeformed oscillator, while preserving the degeneracy associated with rotational symmetry.
}

\vspace{1.3in}

\newpage
\tableofcontents

\newpage

\section{Introduction}
\label{sec_introduction}

The possibility that quantum kinematics may depart from the conventional relation between position and momentum has motivated extensive studies of generalized uncertainty relations and deformed commutation relations. Constructions inspired by quantum gravity have shown how a nonzero minimum position uncertainty may arise from modified position-momentum relations together with suitable representations of the corresponding observables \cite{Maggiore:1993,Kempf:1995,Hossenfelder:2013}. More generally, a modification of the spatial displacement law changes the momentum generators and can therefore affect both localization and dynamics. The construction considered below belongs to this broader class of deformed kinematics, although it does not imply a universal minimum length.

Position-dependent deformations provide a useful setting in which these effects can be studied because spatially varying momentum operators may be related to position-dependent masses and effective configuration-space geometries \cite{Quesne:2004}. Harmonic oscillators with deformed position-momentum relations have also been studied in arbitrary spatial dimension \cite{Chang:2002}. When the system contains a distinguished center, the deformation may depend only on the radius while preserving rotational symmetry, but rotational invariance does not require radial and tangential motion to respond in the same way. This distinction becomes important when a deformation originally formulated for a single coordinate is extended to a system with angular degrees of freedom.

A complementary motivation arises from the nonadditive composition law of generalized entropy \cite{Tsallis:1988}, which leads to deformed algebraic operations and an associated scalar differential calculus \cite{Nivanen:2003,Borges:2004,Borges:2021Corrigenda}. Related ideas have also been considered in quantum-statistical descriptions of interacting systems \cite{Ge:2025Tsallis}. The scalar $q$-calculus used here is associated with this nonadditive composition law and should be distinguished from $q$-oscillator and quantum-group constructions \cite{Biedenharn:1989,Macfarlane:1989}. Applied to a spatial coordinate, it produces a derivative with an affine position-dependent coefficient, which can be interpreted as the local generator of a deformed displacement. The statistical origin of this structure supplies the motivation for the deformation, while its realization as a multidimensional quantum theory requires a separate specification of the observables, measure and dynamics.

Nonadditive displacement laws have previously been implemented through position-dependent momentum operators and effective-mass Hamiltonians \cite{CostaFilho:2011}. Subsequent developments established Hermitian realizations, generalized position and momentum variables, mappings between deformed oscillators and conventional quantum systems, and corresponding uncertainty relations \cite{Mazharimousavi:2012,DaCosta:2014,CostaFilho:2012}. A three-dimensional construction also connected deformed commutators, position-dependent masses and curved spaces using Hermitian momentum operators whose derivative terms share a common positive function of position, with a linear radial function studied in a Coulomb example \cite{Quesne:2004}. More recently, a formulation based on the scalar $q$-calculus and its dual derivative has been related to a Hermitian Hamiltonian with position-dependent mass and an associated effective geometry, with an application to the harmonic oscillator in the weak-deformation regime \cite{Boumali:2026}. These results establish that radial position-dependent momentum operators are not by themselves new. The distinction relevant here is that the scalar construction of \cite{Quesne:2004} rescales radial and tangential momentum directions by the same function, whereas the completion considered below retains the deformed radial derivative while leaving the tangential response undeformed. This choice leads to a different geometric measure, momentum algebra and quantum Hamiltonian.

The central question of the present work is therefore whether the radial derivative and integration measure inherited from the scalar calculus can be embedded into a consistent rotationally covariant quantum theory without requiring the same deformation of the tangential directions. The scalar law alone cannot answer this question because it fixes neither the angular response nor the full configuration-space measure. Our contribution is to specify one direct completion in which these ingredients are fixed together, and then to determine its consequences for the momentum observables, localization, radial dynamics and bound-state spectrum. The construction provides a possible multidimensional extension of the scalar $q$-calculus, with specific additional assumptions whose physical consequences can be calculated and examined.

To implement this completion, we introduce the dimensionless deformation parameter $q$ together with a reference length $\ell_q$. They enter the spatial theory through the positive inverse-length scale $k=(1-q)/\ell_q$, so the branch considered here corresponds to $q<1$. The ordinary spatial theory is recovered as $k$ tends to zero. We use a local rotationally covariant frame to relate the canonical momenta to the momentum components selected by the model, retain the radial derivative inherited from the scalar calculus, and lift its radial measure with the standard spherical measure. Requiring this measure to agree with the geometric volume associated with the frame fixes the tangential frame coefficient to its undeformed value. The resulting model therefore deforms the radial principal direction while preserving the ordinary tangential response.

The same construction defines a spherically symmetric configuration-space geometry. We use $r$ as the areal radius, so that spheres at fixed $r$ retain their standard area, whereas the additive radial coordinate inherited from the scalar calculus coincides with the proper radial distance of the induced geometry. The two radial variables are therefore physically distinct. The geometry has negative and position-dependent curvature away from a singular center and approaches constant negative curvature asymptotically. Rotations about the center remain exact symmetries, while ordinary spatial translations are broken. The distinguished center and the preferred local frame should accordingly be regarded as physical ingredients of the effective model, as would occur for a radial medium, localized source or defect.

On this background we construct the Cartesian frame-momentum components as self-adjoint generators of unitary weighted flows. The canonical momenta conjugate to the Cartesian coordinates remain available and obey the ordinary Heisenberg algebra, whereas the frame momenta describe the mechanical components resolved along the selected local directions. Their mutual commutator contains angular momentum with a radius-dependent coefficient, which prevents all Cartesian frame components from being diagonalized simultaneously as commuting momentum variables. Rotational covariance is nevertheless preserved and allows the dynamics to be organized in angular momentum sectors. This explicit relation between the chosen radial frame and the resulting momentum algebra is one of the main structural results of the construction.

The Robertson inequality gives uncertainty bounds for the individual position and frame-momentum components, and combining these bounds yields a rotationally invariant relation whose correction depends on the mean coordinate radius about the distinguished center, without requiring rotational symmetry of the probability distribution \cite{Robertson:1929}. Position-dependent corrections to uncertainty relations already appear in nonadditive displacement models \cite{CostaFilho:2011}, with subsequent work examining Hermitian momentum operators and their uncertainty properties \cite{DaCosta:2014}. For a scalar deformation linear in radius, applying the same uncertainty inequality to the mixed commutator given in \cite{Quesne:2004} also yields a correction proportional to the mean radius, so the dependence on the mean radius is not unique to the present construction. In the present model, the chosen radial and tangential frame coefficients determine the component uncertainty bounds, which can be combined into a rotationally invariant relation without assuming rotational symmetry of the probability distribution. The mean radius is distinct from the position spread, and the resulting relation does not impose a universal minimum coordinate uncertainty.

The spectral interpretation of the momentum operators requires additional care. Fourier constructions introduced in nonextensive statistics have their own definitions and inversion properties \cite{Umarov:2008,Hilhorst:2010} and do not automatically provide the spectral transform of the quantum momenta considered here. Each self-adjoint Cartesian frame component admits its own spectral resolution, whereas the symmetric radial projection of these components is subject to the standard half-line obstruction. The additive radial coordinate covers only a half-line, for which the corresponding first-order scalar radial momentum has no self-adjoint extension with the same local action \cite{Bonneau:2001}. We therefore formulate radial dynamics through the chosen self-adjoint second-order Hamiltonian, with the admissible behavior at the center determined by its kinetic-energy form.

We define the kinetic energy through the positive quadratic form constructed from the self-adjoint frame-momentum components. This gives a non-negative self-adjoint Hamiltonian and fixes the center condition in the angular sectors where the local radial equation permits more than one realization. The resulting prescription may be compared with the Laplace-Beltrami Hamiltonian associated with the same spatial metric. Both reproduce the same classical kinetic energy, but their quantum expressions differ by a calculable scalar contribution, reflecting the familiar ordering freedom of position-dependent-mass systems \cite{vonRoos:1983,Quesne:2004}. Comparing the two prescriptions therefore allows the dependence of spectral predictions on the quantum kinetic term to be separated from the effects of the underlying radial geometry.

The isotropic harmonic oscillator provides a direct dynamical test because its undeformed spectrum contains degeneracies between states with different radial and orbital quantum numbers in addition to the degeneracy required by rotations. Deformed and position-dependent-mass oscillators have been studied in several settings \cite{Chang:2002,Quesne:2007,Boumali:2026}, while oscillators on nonconstant-curvature spaces can also retain enhanced degeneracies when additional conserved quantities are present \cite{Ballesteros:2011}. Our question is more specific. We keep a potential quadratic in the areal radius and ask how the direct radial completion modifies the shell structure. We obtain the radial differential operator at finite deformation and calculate the leading energy corrections when the oscillator length is much smaller than $k^{-1}$, keeping the radial and orbital quantum numbers fixed. These corrections depend on the two quantum numbers separately and lift degeneracies between states in different orbital angular momentum sectors of the same undeformed oscillator shell, while preserving the degeneracy required by rotations. We verify this behaviour numerically in three dimensions and compare it with the corresponding prediction obtained from the Laplace-Beltrami kinetic prescription.

The same geometric measure also determines the spatial factor entering the leading high-energy semiclassical count of states inside a ball of fixed radius centered at the distinguished point. At small areal radius the enclosed volume has the ordinary Euclidean scaling, whereas at large areal radius its leading power is reduced by one and the volume becomes proportional to the area of the bounding sphere, with the deformation length setting the proportionality scale. At large proper radial distance, the same volume grows exponentially, consistently with the asymptotic negative curvature. We characterize this crossover by an effective radial scaling exponent only as a geometric diagnostic. A hyperscaling-violation exponent in field theory or holography instead refers to the scaling of specified physical observables \cite{Dong:2012se}, so the present volume law does not by itself imply a reduced spatial dimension, an entropy area law or a holographic dual description.

The preferred frame determines the momentum algebra and uncertainty bound, while its geometric measure controls the large-radius volume and the spatial factor in the semiclassical state count. The oscillator spectrum depends additionally on the chosen kinetic-energy prescription and on the confining potential, which is taken to be quadratic in the coordinate radius. These results identify the roles of the spatial construction and the dynamical assumptions, while a microscopic realization would also need to explain the distinguished center and the preferred frame.

The paper is organized as follows. In \cref{q_review} we review the nonadditive composition law, its differential calculus and the Fourier constructions relevant to the present problem. In \cref{sec_radial_kinematics} we introduce the direct radial completion and derive its geometry, Hilbert-space measure and local frame-momentum expressions. In \cref{sec:momenta_uncertainty} we construct the self-adjoint frame generators, determine their algebra and radial spectral properties, define the frame Hamiltonian and analyze the uncertainty bound, volume scaling and harmonic-oscillator spectrum. Finally, we discuss the main results, their physical implications and the limitations of the model in \cref{dis}, while \cref{app_weak_domain_form_core} establishes the approximation result that connects the global kinetic-energy form with the local differential expression and justifies the conditions imposed at the center.

\section{Review of nonadditive algebra and calculus}\label{q_review}

Before introducing the deformed quantum kinematics used in the present work, it is useful to review the algebraic structures that arise in nonextensive statistical mechanics. The nonextensive formalism gives rise to several mutually related algebraic structures. Among these, relevant to the present construction are the $q$-additive or pseudo-additive composition law associated with the Tsallis entropy for statistically independent subsystems \cite{Tsallis:1988}, the related $q$-deformed sum and algebra \cite{Nivanen:2003}, and a compatible choice of $q$-differential calculus \cite{Borges:2004}. These structures provide the algebraic motivation for the deformation introduced in our scenario.  However, they do not uniquely determine how this deformation should be implemented in quantum mechanics, and the resulting construction, therefore, must be specified and analyzed independently.

\subsection{Tsallis entropy and the $q$-composition law}
\label{tsallis_q_sum}
For a discrete probability distribution $\{p_i\}$, satisfying $p_i\geq0$ and $\sum_{i=1}^W p_i=1$, the Tsallis entropy is defined by \cite{Tsallis:1988}
\begin{align}
	S_q
	=
	{k_B \over q-1}
	\left(
	1-\sum_{i=1}^W p_i^q
	\right)~,
	\label{Tsallis_entropy_review}
\end{align}
where $q\in\mathbb{R}$ is the nonextensivity parameter, or deformation parameter, while $W\in\mathbb{N}$ denotes the number of accessible microstates. One can note that the ordinary Boltzmann-Gibbs entropy can be recovered continuously in the limit
$q\to1$,
\begin{align}
	\lim_{q\to1}S_q
	=
	-k_B\sum_{i=1}^W p_i\log p_i~.
\end{align}
The nonadditive character of the Tsallis entropy is manifest in its composition rule for two statistically independent systems $A$ and $B$, whose combined probability distribution factorizes as $p_{ij}^{A \cup B}=p_i^A p_j^B$. Consequently, from the definition of the Tsallis entropy, one obtains
\begin{align}
	S_q(A\cup B)
	=
	S_q(A)+S_q(B)
	+
	\frac{1-q}{k_B}S_q(A)S_q(B)~.
	\label{Tsallis_composition}
\end{align}
This relation is the pseudo-additive, or $q$-additive, composition law of the Tsallis entropy, while the ordinary additive law is recovered in the limit $q\to1$. The algebraic structure suggested by \cref{Tsallis_composition} is encoded in the $q$-sum \cite{Nivanen:2003,Borges:2004,Umarov:2008},
\begin{align}
	u\oplus_q v
	=
	u+v+(1-q)\,u\,v~,
	\label{q_sum}
\end{align}
where $u$ and $v$ are dimensionless real variables. In terms of this $q$-addition operation, the entropy composition law in \cref{Tsallis_composition} can be expressed as
\begin{align}
	s_q(A\cup B)
	=
	s_q(A)
	\oplus_q
	s_q(B)~,
\end{align}
with $s_q=S_q/ k_B$. One can note that the $q$-sum satisfies the commutative and associative laws
\begin{align}
	u\oplus_q v
	&=
	v\oplus_q u~,
	\notag\\
	(u\oplus_q v)\oplus_q w
	&=
	u\oplus_q(v\oplus_q w)~.
\end{align}
The neutral element is the ordinary zero,
\begin{align}
	u\oplus_q0=u~.
\end{align}
The additive inverse with respect to $\oplus_q$ is
\begin{align}
	\ominus_q u
	=
	-{u\over1+(1-q)u}~,
	\label{q_inverse}
\end{align}
provided $1+(1-q)u\neq0$. The corresponding $q$-difference is
\begin{align}
	u\ominus_q v
	&=
	u\oplus_q(\ominus_q v)
	=
	{u-v\over1+(1-q)v}~.
	\label{q_diff}
\end{align}
Furthermore a useful identity is
\begin{align}
	1+(1-q)(u\oplus_q v)
	=
	\left[
	1+(1-q)u
	\right]
	\left[
	1+(1-q)v
	\right].
	\label{q_sum_multiplicative_identity}
\end{align}
It shows that the nonlinear composition law can be mapped to ordinary addition by an appropriate logarithmic coordinate. This observation will be useful for the later part of our analysis.

\subsection{$q$-logarithm, $q$-exponential and deformed differential calculus}
\label{q_log_exp_calculus_review} 
The $q$-logarithm and the $q$-exponential are defined as \cite{Tsallis:1988,Umarov:2008,Borges:2004}
\begin{align}
	\log_q z
	&=
	\frac{z^{1-q}-1}{1-q},
	\qquad z>0~,\notag\\
	e_q(u)
	&=
	\left[
	1+(1-q)u
	\right]^{\frac{1}{1-q}}~,
	\label{q_exp_review}
\end{align}
where, for real $u$, the latter is defined on the domain $1+(1-q)u>0$. We would like to emphasize here that throughout this work we use this real branch defined on the domain where the corresponding deformation factor is positive. These functions are mutually inverse,
\begin{align}
	\log_q\left(e_q(u)\right)
	=
	u~,
	\qquad
	e_q\left(\log_q z\right)
	=
	z~,
\end{align}
and reduce to the ordinary logarithm and exponential in the undeformed limit,
\begin{align}
	\lim_{q\to1}\log_q z
	&=
	\log z~,
	&
	\lim_{q\to1}e_q(u)
	&=
	e^u~.
\end{align}
The $q$-logarithm maps ordinary multiplication into the deformed sum,
\begin{align}
	\log_q(zw)
	=
	\log_q z
	\oplus_q
	\log_q w~,
	\label{q_log_product}
\end{align}
or, equivalently,
\begin{align}
	e_q(u)e_q(v)
	=
	e_q(u\oplus_q v)~.
	\label{q_exp_product}
\end{align}
These identities express the compatibility of the $q$-exponential and $q$-logarithm with the non-additive composition law.

A differential calculus adapted to the $q$-sum was introduced in \cite{Borges:2004} where, in terms of the $q$-difference, the corresponding $q$-derivative is defined as
\begin{align}
	D_q f(u)
	=
	\lim_{v\to u}
	\frac{f(u)-f(v)}
	{u\ominus_q v}~.
	\label{Borges_derivative_definition}
\end{align}
Utilizing $q$-difference, described in \cref{q_diff}, one obtains
\begin{align}
	D_q f(u)
	=
	\left[
	1+(1-q)u
	\right]
	\frac{df(u)}{du}~.
	\label{Borges_derivative_def}
\end{align}
Now the ordinary derivative of the $q$-exponential satisfies
\begin{align}
	\frac{d}{du}e_q(u)
	=
	\left[e_q(u)\right]^q.
\end{align}
Together with
\begin{align}
	1+(1-q)u
	=
	\left[e_q(u)\right]^{1-q},
\end{align}
this implies
\begin{align}
	D_q e_q(u)
	=
	e_q(u).
	\label{q_exponential_eigenfunction}
\end{align}
Thus, just as the ordinary exponential is an eigenfunction of the ordinary derivative, the $q$-exponential is an eigenfunction of the  $q$-derivative.

The relation between the  $q$-derivative and the additive coordinate of the $q$-translation group can be made explicit by introducing
\begin{align}
	\tau_q(u)
	=
	{1\over1-q}
	\log\left[
	1+(1-q)u
	\right]~.
	\label{q_group_coordinate}
\end{align}
Utilizing the multiplicative identity, given in \cref{q_sum_multiplicative_identity}, associated with the $q$-sum, one finds
\begin{align}
	\tau_q(u\oplus_qv)
	=
	\tau_q(u)+\tau_q(v)~.
	\label{q_group_coordinate_addition}
\end{align}
This is why $\tau_q$ is called the additive group coordinate or group logarithm of the $q$-sum. Moreover,
\begin{align}
	{d\tau_q\over du}
	=
	{1\over1+(1-q)u}~,\label{Borges_derivative}
\end{align}
so that one obtains
\begin{align}
	D_q
	=
	{\partial\over\partial\tau_q}~.
	\label{Borges_group_derivative}
\end{align}
The $q$-derivative is therefore equivalent to the ordinary derivative with respect to the additive coordinate $\tau_q$. This relation is a direct consequence of the underlying group structure. Indeed, \cref{q_group_coordinate_addition} shows that a $q$-translation $u\to u\oplus_q v$ acts as an ordinary translation
\begin{align}
	\tau_q(u)
	\longrightarrow
	\tau_q(u)+\tau_q(v)
\end{align}
in the additive coordinate. Consequently, when $\tau_q(v)$ is considered to be infinitesimal, the corresponding translation generator is given by \cref{Borges_group_derivative}. The same additive coordinate also determines the invariant integration measure from \cref{Borges_derivative} as
\begin{align}
	d_qu
	\equiv
	d\tau_q
	=
	{du\over1+(1-q)u}~.
	\label{q_invariant_measure}
\end{align}
Accordingly, the $q$-sum, the $q$-derivative and the measure $d_qu$ are not independent deformations, but rather different realizations of the same underlying translation structure.

Up to this point, $D_q$ is an algebraic operator defined within the $q$-differential calculus. However, in the quantum-mechanical construction developed in this work, we adopt this algebraic structure and investigate its spatial realization by interpreting the associated derivative as the infinitesimal generator of deformed translations. This constitutes the additional physical input that connects the nonextensive algebra to the deformed quantum kinematics considered in the present work.

\subsection{$q$-Fourier analysis}
\label{review_q_fourier}
Several inequivalent constructions are referred to as a $q$-Fourier transformation in the literature \cite{Olshanetsky:1999,Umarov:2008,Nelson:2008,Kalogeropoulos:2018}. It is therefore essential to distinguish between the $q$-Fourier transforms formulated for nonextensive probability distributions and the transformation between the position and momentum representations required in quantum mechanics.

The ordinary Fourier transformation,
\begin{align}
	\widetilde f(\xi)
	=
	{1\over\sqrt{2\pi}}
	\int_{\mathbb R}du\,
	e^{-i\xi u}f(u),
	\label{ordinary_Fourier_review}
\end{align}
is built from the unitary characters of the additive translation group. The kernel satisfies
\begin{align}
	e^{i\xi(u+v)}
	=
	e^{i\xi u}e^{i\xi v},
	\label{ordinary_character}
\end{align}
and diagonalizes the infinitesimal translation generator,
\begin{align}
	-i{\partial\over\partial u}e^{i\xi u}
	=
	\xi e^{i\xi u}.
	\label{ordinary_eigenvalue}
\end{align}
Linearity, orthogonality, completeness and the Plancherel theorem all follow from this spectral structure.

In nonextensive probability theory, \cite{Umarov:2008} introduced a nonlinear $q$-Fourier transform in connection with the $q$-generalized central-limit theorem which takes the schematic form
\begin{align}
	{\cal F}^{\rm UTS}_q[f](\xi)
	=
	\int_{\mathbb R}du\,
	f(u)\,
	e_q\left(
	i\xi u[f(u)]^{q-1}
	\right)~,
	\label{UTS_transform_review}
\end{align}
for an appropriate class of non-negative functions $f$. The detailed definition may equivalently be expressed using the $q$-product. The kernel in \cref{UTS_transform_review} depends explicitly on the function being transformed and consequently, the map is nonlinear as
\begin{align}
	{\cal F}^{\rm UTS}_q[a\,f+b\,g]
	\neq
	a\,{\cal F}^{\rm UTS}_q[f]
	+
	b\,{\cal F}^{\rm UTS}_q[g]
\end{align}
for generic $f$, $g$, $a$ and $b$. Such a transformation is not directly suitable as a quantum mechanical change of representation. A position and momentum transformation acting on wavefunctions should be linear, should preserve the Hilbert-space inner product and should be generated by the generalized eigenfunctions associated with the spectral resolution of a specified self-adjoint momentum operator. 

A second possibility is the formal replacement
\begin{align}
	e^{i\xi u}
	\longrightarrow
	e_q(i\xi u)~.
	\label{formal_q_kernel}
\end{align}
Although \cref{formal_q_kernel} has the correct undeformed limit, this condition alone is insufficient. A Fourier kernel associated with translations must diagonalize the corresponding translation generator, define a unitary or at least invertible transformation on an appropriate function space, satisfy the character property of the deformed translation group, and admit consistent orthogonality and completeness relations with respect to the invariant measure. The direct complex $q$-exponential generally fails these requirements. For real $q$, $\xi$ and $u$, the formal complex $q$-exponential has modulus
\begin{align}
	\left|e_q(i\xi u)\right|
	=
	\left[1+(1-q)^2\xi^2u^2\right]^{1/[2(1-q)]} ,
	\label{q_exp_modulus}
\end{align}
on the principal complex branch. This should be distinguished from the real branch used in \cref{q_log_exp_calculus_review}, where the argument of $e_q$ was real and the deformation factor was required to be positive. Thus, unlike the ordinary Fourier kernel, the direct complex $q$-exponential contains a nontrivial amplitude in addition to the phase. This observation is important in quantum mechanics where a plane-wave kernel should be a pure phase, such that the corresponding translation operator is unitary and preserves the norm of the state. The appropriate unit-modulus plane-wave mode associated with the additive coordinate given in \cref{q_group_coordinate} may instead be written as
\begin{align}
\chi_\xi(u)
=
\exp\left(i\xi\tau_q(u)\right)
=
\left[1+(1-q)u\right]^{i\xi/(1-q)}~,
\label{q_unitary_character_review}
\end{align}
which satisfies the eigenvalue equation $D_q\chi_\xi=i\xi\chi_\xi$ on the real domain $1+(1-q)u>0$. The consequences of using this plane-wave eigenfunction will be discussed in detail in \cref{subsec_radial_characters_obstruction}. Therefore, a formal replacement $e^{i\xi u}\to e_q(i\xi u)$ does not, by itself, define a position and momentum transform with the usual unitarity and orthogonality properties.

The discussion above shows that the algebraic structures of nonextensive statistical mechanics do not automatically define a quantum theory. In particular, a consistent quantum-mechanical construction requires a Hilbert-space measure, densely defined symmetric operators, specified self-adjoint realizations of the observables, and an appropriate spectral problem. In the following section we formulate a radial model while keeping the additional multidimensional and operator-theoretic assumptions explicit.

\section{A $q$-inspired radial quantum kinematics}
\label{sec_radial_kinematics}

As reviewed in \cref{q_review}, the nonextensive formalism naturally gives rise to a scalar composition law together with the associated $q$-differential calculus. These structures are, however, algebraic and one-dimensional and, in their standard form, are formulated in terms of dimensionless variables. Their application to quantum mechanics on physical space therefore requires additional physical input. First, a reference length scale must be introduced in order to convert the dimensionless deformation into a dimensionful parameter appropriate for physical space. Second, a consistent multi-dimensional phase-space structure must be specified. In particular, the scalar $q$-sum and the $q$-derivative do not, by themselves, determine the multi-dimensional position-momentum algebra, the Hilbert-space measure, or the domains and self-adjoint realizations of the resulting observables.

\subsection{Dimensionful radial deformation}
\label{measure_deformation}

We introduce a positive reference length $\ell_q$ and define $k=(1-q)/\ell_q$, which has dimensions of inverse length, restricting the following analysis to $k>0$ and hence $q<1$. Since the kinematics depends only on $k$, the parameters $q$ and $\ell_q$ cannot be determined separately within the present model unless one of them is fixed by independent physical input. We refer to the construction as $q$-inspired because its radial ingredients are motivated by the scalar $q$-calculus, while their extension to a rotationally covariant quantum theory requires additional assumptions that will be specified in the following subsection.

Writing the dimensionless variables in \cref{q_sum} as $u=x/\ell_q$ and $v=y/\ell_q$, and multiplying their composition by $\ell_q$, gives the corresponding law for quantities with dimensions of length,
\begin{align}
	x\oplus_k y
	=
	x+y+kxy~.
	\label{dimensionful_q_sum}
\end{align}
On the admissible interval $\mathcal I_k=(-k^{-1},\infty)$, this composition can be expressed as ordinary addition through a change of coordinate. The identity
\begin{align}
	1+k(x\oplus_k y)=(1+kx)(1+ky)
	\label{full_line_multiplicative_identity}
\end{align}
implies that the coordinate
\begin{align}
	\tau_k(x)=\frac{1}{k}\log(1+kx)
	\label{full_line_additive_coordinate}
\end{align}
satisfies
\begin{align}
	\tau_k(x\oplus_k y)=\tau_k(x)+\tau_k(y)~.
	\label{full_line_additive_law}
\end{align}
As $x$ ranges over $\mathcal I_k$, the coordinate $\tau_k$ covers the entire real line, on which the composition law therefore takes the usual additive form.

We consider systems with a natural center, such as a particle in a central potential, and introduce the scalar deformation while preserving their rotational symmetry. Deforming a single Cartesian coordinate, or applying independent deformations to all Cartesian components, would select preferred axes. We therefore use the coordinate radius $r=|\mathbf{x}|$ as the variable carrying the deformation, since it is invariant under rotations about the selected center. Applying \cref{dimensionful_q_sum} to non-negative radial scales gives
\begin{align}
	r\oplus_k a=r+a+kra~,
	\qquad r,a\in[0,\infty)~.
	\label{radial_q_sum}
\end{align}
This composition is associative and commutative, with identity zero, but a positive radius has no inverse within the physical radial domain because the solution of $r\oplus_k a=0$ is $a=-r/(1+kr)$, which is negative for $r>0$. The radial composition law does not determine how motion along the spheres at fixed radius should be modified. To isolate the consequences of deforming radial motion, the completion introduced below retains the ordinary tangential response as an additional assumption, allowing us to study the scalar deformation in a system with angular degrees of freedom.

The identity described in \cref{full_line_multiplicative_identity} for the radial coordinate
\begin{align}
	1+k(r\oplus_k a)=(1+kr)(1+ka)
	\label{radial_multiplicative_identity}
\end{align}
allows us to use the same change of coordinate,
\begin{align}
	\tau_k(r)=\frac{1}{k}\log(1+kr)~,
	\qquad
	\frac{d\tau_k}{dr}=\frac{1}{1+kr}~,
	\label{radial_additive_coordinate}
\end{align}
for which
\begin{align}
	\tau_k(r\oplus_k a)=\tau_k(r)+\tau_k(a)~.
	\label{radial_additive_law}
\end{align}
The restriction to non-negative radii now gives
\begin{align}
	r\in[0,\infty)
	\quad\Longleftrightarrow\quad
	\tau_k\in[0,\infty)~,
	\label{radial_half_line_map}
\end{align}
so the additive coordinate covers only a half-line, whose endpoint at $\tau_k=0$ is essential for the radial momentum analysis in \cref{subsec_radial_characters_obstruction}.

For a fixed displacement parameter $a\geq0$, the transformation $T_a(r)=r\oplus_k a$ satisfies
\begin{align}
	\frac{dT_a(r)}{1+kT_a(r)}
	=
	\frac{dr}{1+kr}~.
	\label{radial_measure_invariance}
\end{align}
Although this relation preserves the differential measure, for $a>0$ the image of the radial domain is $T_a([0,\infty))=[a,\infty)$, so the inverse exists only on this image and not on the entire physical half-line. Consequently, \cref{radial_measure_invariance} does not establish a global radial translation group acting on the physical domain.

The radial measure and derivative associated with the additive coordinate are
\begin{align}
	d\tau_k=\frac{dr}{J(r)}~,
	\qquad
	D_k^{(r)}=J(r)\partial_r=\partial_{\tau_k}~,
	\qquad
	J(r)\equiv1+kr~.
	\label{radial_Borges_derivative}
\end{align}
These radial ingredients inherited from the scalar $q$-calculus determine the coefficient of the radial derivative, but do not specify the derivatives along angular directions or the full configuration-space measure. In particular, the volume factor associated with the spheres at fixed $r$ must be included when constructing the multidimensional theory, whose additional assumptions are given in the following subsection.

Since $r=|\mathbf{x}|$ is measured from a selected origin, the construction below preserves rotations about that origin while breaking ordinary spatial translation invariance. Its physical interpretation is therefore that of an effective kinematics relative to a distinguished center, which may represent a fixed source or defect.

\subsection{Direct rotationally covariant radial completion}
\label{tensorial_phase_space_algebra}

To specify the additional input required for the multidimensional construction, we start from canonical phase-space variables $(x^i,p_j)$,
\begin{align}
	\{x^i,p_j\}=\delta^i{}_j~,
	\qquad
	\{x^i,x^j\}=0~,
	\qquad
	\{p_i,p_j\}=0~,
	\label{canonical_phase_space}
\end{align}
and introduce the momentum variables
\begin{align}
	P_i=E_i{}^a(x)p_a~,
	\label{physical_momentum}
\end{align}
where the real invertible frame $E_i{}^a$ determines how the canonical momentum is resolved into the selected momentum components. Their position-momentum bracket is
\begin{align}
	\{x^j,P_i\}=E_i{}^j(x)~.
	\label{classical_deformed_mixed_bracket}
\end{align}
The modified bracket therefore follows from the choice of $P_i$, while the canonical variables continue to satisfy \cref{canonical_phase_space}. In the quantum construction, the derivative term associated with $P_i$ will be $-i\hbar E_i{}^a\partial_a$, with the accompanying multiplication term and operator domain specified after choosing the Hilbert-space
measure.

For $\mathbf{x}\neq0$, we define the radial unit vector and the radial and transverse projections by
\begin{align}
	n^i=\frac{x^i}{r}~,
	\qquad
	(\mathcal P_r)^i{}_j=n^i n_j~,
	\qquad
	(\mathcal P_\perp)^i{}_j=\delta^i{}_j-n^i n_j~.
	\label{radial_tangential_projectors}
\end{align}
We consider real, symmetric, positive-definite frames that transform covariantly under rotations and reflections about the selected origin,
\begin{align}
	E(Rx)=R E(x)R^{\mathsf T}~,
	\qquad R\in O(d)~.
	\label{frame_Od_equivariance}
\end{align}
Within this class, symmetry under transformations that leave the radial direction fixed requires all transverse directions to have the same coefficient and excludes mixing between radial and transverse components. The frame therefore takes the form
\begin{align}
	E_i{}^a
	=
	A_\perp(r)(\mathcal P_\perp)_i{}^a
	+
	A_r(r)(\mathcal P_r)_i{}^a~,
	\qquad
	A_r(r)>0~,
	\quad
	A_\perp(r)>0~.
	\label{general_rot_covariant_frame}
\end{align}
Its radial projection and determinant are
\begin{align}
	n^iE_i{}^a\partial_a
	&=A_r(r)\partial_r~,
	\label{radial_projection_general_frame}\\
	\det E
	&=A_r(r)A_\perp(r)^{d-1}~.
	\label{det_E_general}
\end{align}
Matching the radial derivative in \cref{radial_Borges_derivative} fixes
\begin{align}
	A_r(r)=J(r)=1+kr~,
	\label{radial_eigenvalue_fixed}
\end{align}
while the transverse coefficient remains undetermined. Consequently,
\begin{align}
	\det E=J(r)A_\perp(r)^{d-1}
	\label{det_not_fixed_by_scalar_input}
\end{align}
is not fixed by the scalar radial $q$-calculus alone.

We select the model by adopting the configuration-space measure
\begin{align}
	d\mu_k
	=
	\frac{r^{d-1}dr\,d\Omega_{d-1}}{J(r)}
	=
	\frac{d^dx}{J(r)}~,
	\label{radial_completion_measure_choice}
\end{align}
which combines the radial differential measure in \cref{radial_Borges_derivative} with the usual spherical volume factor $r^{d-1}d\Omega_{d-1}$. We further require this measure to coincide with the geometric volume associated with the frame. Defining the inverse metric by
\begin{align}
	g^{ab}=\delta^{ij}E_i{}^aE_j{}^b~,
	\qquad
	\sqrt{g}=\frac{1}{\det E}~,
	\qquad
	g\equiv\det(g_{ab})~,
	\label{radial_completion_frame_volume}
\end{align}
the equality $\sqrt{g}\,d^dx=d\mu_k$ requires $\det E=J$. Together with the radial derivative condition, this gives
\begin{align}
	A_r(r)=J(r)~,
	\qquad
	\det E=J(r)~.
	\label{direct_radial_requirements}
\end{align}
These conditions define the direct radial completion used below. Substituting them into \cref{det_E_general} gives
\begin{align}
	A_\perp(r)^{d-1}=1~,
	\label{direct_radial_tangential_choice}
\end{align}
so positivity of the frame and $d\geq2$ determine
\begin{align}
	A_r(r)=J(r)~,
	\qquad
	A_\perp(r)=1~.
	\label{direct_radial_completion}
\end{align}
The radial coefficient is therefore deformed while the transverse coefficient retains its undeformed value. This result follows from the combined radial derivative and volume prescriptions, which constitute additional multidimensional assumptions beyond the scalar $q$-calculus.

For $r>0$, the resulting frame is
\begin{align}
	E_i{}^a
	&=
	(\mathcal P_\perp)_i{}^a
	+J(r)(\mathcal P_r)_i{}^a
	=
	\delta_i{}^a+kr\,n_i n^a
	=
	\delta_i{}^a+k\frac{x_i x^a}{r}~,
	\label{direct_radial_frame}\\
	\det E&=J(r)~.
	\label{direct_radial_determinant}
\end{align}
These coefficients extend continuously to the center, where we define $E_i{}^a(0)=\delta_i{}^a$, consistently with the global construction developed below. Its unequal radial and transverse coefficients are compatible with rotational symmetry because the radial direction is determined by the position relative to the selected center, rather than by a fixed spatial axis.

For comparison, a scalar frame deformation $E_i{}^a=f(r)\delta_i{}^a$ assigns equal radial and transverse coefficients, $A_r=A_\perp=f$. Reproducing the radial derivative $J\partial_r$ would then require $f=J$, giving $\det E=J^d$. Alternatively, matching the prescribed volume measure requires $f=J^{1/d}$, whose radial derivative is $J^{1/d}\partial_r$. Thus, for $d>1$ and $k\neq0$, equal radial and transverse coefficients cannot reproduce both the radial derivative and the volume measure selected in \cref{direct_radial_requirements}. Both choices preserve rotations about the origin, but they assign different frame coefficients to the same position coordinates. Here we adopt the direct radial completion in \cref{direct_radial_completion}.

The same determinant enters the classical phase-space measure. At fixed position, the transformation $P_i=E_i{}^a p_a$ gives $d^dP=(\det E)d^dp$, and retaining $x$ as the configuration variable therefore yields
\begin{align}
	\frac{d^dx\,d^dp}{(2\pi\hbar)^d}
	=
	\frac{d^dx\,d^dP}{(2\pi\hbar)^d\det E(x)}~.
	\label{phase_space_measure_general_frame}
\end{align}
For the selected frame this becomes
\begin{align}
	d\Gamma_k
	=
	\frac{r^{d-1}dr\,d\Omega_{d-1}\,d^dP}
	{(2\pi\hbar)^dJ(r)}~.
	\label{phase_space_measure_direct_radial}
\end{align}
The above equation expresses the canonical Liouville measure in the variables $(x,P)$, and its configuration-space factor agrees with the volume measure adopted in \cref{radial_completion_measure_choice}. This agreement provides a consistency check between the classical momentum transformation and the geometric prescription, while the quantum Hilbert space and its inner product are specified in the following subsection.

\subsection{Geometric interpretation and Hilbert-space measure}
\label{subsec_geometric_interpretation_measure}

For the frame selected in \cref{direct_radial_frame}, the standard relation between an orthonormal frame and the metric gives the inverse metric
\begin{align}
	g^{ab}
	=
	\delta^{ij}E_i{}^aE_j{}^b
	=
	(\mathcal P_\perp)^{ab}
	+
	J(r)^2(\mathcal P_r)^{ab}~.
	\label{inverse_metric_from_frame}
\end{align}
The corresponding metric is
\begin{align}
	g_{ab}
	=
	(\mathcal P_\perp)_{ab}
	+
	J(r)^{-2}(\mathcal P_r)_{ab}~,
	\label{metric_from_frame}
\end{align}
with line element
\begin{align}
	ds^2
	=
	\frac{dr^2}{J(r)^2}
	+
	r^2d\Omega_{d-1}^2~.
	\label{direct_radial_line_element}
\end{align}
A sphere at fixed $r$ has the usual area $\Omega_{d-1}r^{d-1}$, where $\Omega_{d-1}$ denotes the area of the unit $(d-1)$-sphere, while the radial distance from the center measured using the metric is
\begin{align}
	\tau_k(r)
	=
	\int_0^r\frac{ds}{J(s)}
	=
	\frac{1}{k}\log(1+kr)~,
	\label{proper_radial_distance}
\end{align}
which gives a geometric interpretation to the additive coordinate introduced in \cref{radial_additive_coordinate}. In this coordinate, the metric becomes
\begin{align}
	ds^2
	=
	d\tau_k^2+r(\tau_k)^2d\Omega_{d-1}^2~,
	\qquad
	r(\tau_k)=\frac{e^{k\tau_k}-1}{k}~.
	\label{tau_metric_warped_form}
\end{align}
This distinction is relevant to the uncertainty bound in \cref{radial_scale_dependent_gup}, which involves the mean coordinate radius $\langle r\rangle$, while the mean radial distance measured using the metric is $\langle\tau_k(r)\rangle$.

The volume element associated with \cref{direct_radial_line_element} is
\begin{align}
	\sqrt g\,d^dx
	=
	\frac{d^dx}{J(r)}
	=
	\frac{r^{d-1}dr\,d\Omega_{d-1}}{J(r)}~,
	\label{direct_radial_geometric_volume}
\end{align}
in agreement with the measure prescribed in the preceding subsection. The construction therefore admits a description in terms of a particle on a spherically symmetric spatial geometry, with momentum components defined using the selected radial frame.

Although the metric determines local distances and angles, it does not uniquely specify the orthonormal frame in which momentum components are resolved. Frames related by position-dependent rotations describe the same geometry, while the symmetric quantum momentum operators associated with them also contain derivatives of the rotation coefficients. In the present construction, \cref{direct_radial_frame} specifies the momentum components whose operators and uncertainties will be studied. Assigning a preferred role to these components in the quantum Hamiltonian requires a further prescription for the kinetic operator, whose relation to the Laplace-Beltrami prescription is discussed below.

The frame and metric extend continuously to the center, with $E_i{}^a\to\delta_i{}^a$ and $g_{ab}\to\delta_{ab}$ as $r\to0$, but the derivatives of $x_i x^a/r$ do not have a direction-independent limit there. The resulting geometry is singular, as shown by the scalar curvature
\begin{align}
	\mathcal R
	=
	-(d-1)\left[
	\frac{2k(d-1)}{r}+dk^2
	\right]~,
	\qquad r>0~.
	\label{direct_radial_scalar_curvature}
\end{align}
For $k>0$ and $d\geq2$, the curvature diverges at the center, whereas at large radius it approaches $-d(d-1)k^2$. Correspondingly, $r(\tau_k)\sim e^{k\tau_k}/k$ at large proper distance, and the geometry approaches negative constant curvature with length scale $k^{-1}$, although its curvature remains position dependent at finite radius. The center is included in the configuration space and has zero volume measure, but this does not remove its possible influence on the admissible behavior of wavefunctions. The endpoint conditions required to define the radial Hamiltonian are therefore treated separately below.

For scalar wavefunctions, we adopt the Hilbert space defined by the geometric volume measure,
\begin{align}
	\mathcal H_k
	&=
	L^2(\mathbb R^d,d\mu_k)~,
	\qquad
	d\mu_k(x)=w_k(x)d^dx~,
	\qquad
	w_k(x)=\frac{1}{J(r)}~,
	\label{direct_radial_hilbert_space}\\
	\langle\psi,\phi\rangle_k
	&=
	\int_{\mathbb R^d}d\mu_k\,\psi^*(x)\phi(x)
	=
	\int_{\mathbb R^d}\frac{d^dx}{J(r)}
	\,\psi^*(x)\phi(x)~.
	\label{direct_radial_inner_product}
\end{align}
This inner product specifies the normalization of states and the measure with respect to which the momentum operators must be symmetric, while their differential expressions and self-adjoint domains require the additional construction developed below.

\subsection{Canonical and local-frame momenta}
\label{subsec_canonical_local_frame_momenta}

We first define the local differential operators on the common domain
\begin{align}
	\mathcal D_0
	=
	C_c^\infty(\mathbb R^d\setminus\{0\})~,
	\label{punctured_common_test_domain}
\end{align}
whose wavefunctions are smooth and have compact support away from the center. Since $w_k$ is locally bounded and positive, this domain is dense in $\mathcal H_k$, and its support restriction allows integration by parts without boundary contributions. This restriction does not remove the center from the configuration space or determine the boundary conditions of the closed operators, whose self-adjoint realizations are specified separately below.

To identify the canonical momenta conjugate to the Cartesian coordinates, we absorb the position-dependent measure into the wavefunction through the unitary map
\begin{align}
	W\colon\mathcal H_k
	\longrightarrow
	L^2(\mathbb R^d,d^dx)~,
	\qquad
	(W\psi)(x)
	=
	w_k(x)^{1/2}\psi(x)~.
	\label{density_unitary_map_canonical}
\end{align}
This transformation preserves the inner product and allows the ordinary Cartesian momentum operators to be represented on $\mathcal H_k$. For each direction, we define
\begin{align}
	H_i^1(\mathbb R^d)
	=
	\left\{
	f\in L^2(\mathbb R^d,d^dx)
	\,\middle|\,
	\partial_i f\in L^2(\mathbb R^d,d^dx)
	\right\}~,
	\label{directional_sobolev_domain}
\end{align}
where the derivative is understood in the sense of distributions. The canonical momenta and their domains are then
\begin{align}
	\widehat p_i^{\rm can}
	&=
	W^{-1}(-i\hbar\partial_i)W~,
	&
	\mathcal D(\widehat p_i^{\rm can})
	&=
	W^{-1}H_i^1(\mathbb R^d)~.
	\label{canonical_momentum_domain}
\end{align}
Each component is self-adjoint, and the family strongly commutes because the same unitary transformation relates all components to the standard Cartesian momenta \cite{ReedSimonII}. Their common domain is
\begin{align}
	\bigcap_{i=1}^d
	\mathcal D(\widehat p_i^{\rm can})
	=
	W^{-1}H^1(\mathbb R^d)~,
	\label{canonical_momentum_common_domain}
\end{align}
where $H^1(\mathbb R^d)$ requires all first derivatives to be square integrable.

On $\mathcal D_0$, the canonical momenta have the differential expressions
\begin{align}
	\widehat p_i^{\rm can}
	&=
	-i\hbar
	\left[
	\partial_i
	+
	\frac{1}{2}\partial_i\log w_k
	\right]
	\nonumber\\
	&=
	-i\hbar
	\left[
	\partial_i
	-
	\frac{k}{2J(r)}n_i
	\right]~,
	\label{canonical_quantum_momenta}
\end{align}
and satisfy
\begin{align}
	[x^j,\widehat p_i^{\rm can}]
	&=
	i\hbar\delta_i{}^j~,
	&
	[\widehat p_i^{\rm can},\widehat p_j^{\rm can}]
	&=
	0~.
	\label{canonical_quantum_heisenberg_algebra}
\end{align}
The weighted measure changes their differential expressions while preserving the canonical commutators. Their existence does not imply translation invariance of the radial geometry or of a Hamiltonian constructed on it, since a translation generated by a canonical momentum need not be a symmetry of the dynamics.

For the momenta associated with the selected radial frame, we adopt the symmetric ordering of the classical expression $P_i=E_i{}^ap_a$,
\begin{align}
	P_{i,0}
	=
	\frac{1}{2}
	\left(
	E_i{}^a\widehat p_a^{\rm can}
	+
	\widehat p_a^{\rm can}E_i{}^a
	\right)
	\label{frame_momentum_symmetric_from_canonical}
\end{align}
on $\mathcal D_0$. Substituting \cref{canonical_quantum_momenta} gives
\begin{align}
	P_{i,0}
	=
	-i\hbar
	\left[
	E_i{}^a\partial_a
	+
	\frac{1}{2w_k}\partial_a\left(w_kE_i{}^a\right)
	\right]~,
	\label{frame_momentum_weighted_divergence}
\end{align}
which displays how the symmetric ordering incorporates both the spatial variation of the frame and the Hilbert-space weight. This expression defines a symmetric operator on $\mathcal D_0$, while its self-adjoint realization requires the subsequent analysis. The canonical momenta $\widehat p_i^{\rm can}$ are therefore distinct from the symmetric frame operators $P_{i,0}$, whose self-adjoint realizations are constructed in \cref{subsec_self_adjoint_momentum_generators}.

To identify the mechanical meaning of the frame momenta, let $\theta^i=\theta^i{}_a dx^a$ denote the one-forms dual to the frame vectors $E_j{}^a\partial_a$, so that
\begin{align}
	\theta^i{}_aE_j{}^a
	&=
	\delta^i{}_j~,
	&
	g_{ab}
	&=
	\delta_{ij}\theta^i{}_a\theta^j{}_b~.
	\label{dual_coframe_relations}
\end{align}
For the kinetic Lagrangian
\begin{align}
	L
	=
	\frac{m}{2}g_{ab}\dot x^a\dot x^b~,
	\label{radial_particle_lagrangian}
\end{align}
the canonical momentum is $p_a=mg_{ab}\dot x^b$, and its components in the local orthonormal frame are
\begin{align}
	P_i
	=
	E_i{}^ap_a
	=
	m\,\theta_{ia}\dot x^a~,
	\label{classical_local_frame_momentum}
\end{align}
where $\theta_{ia}=\delta_{ij}\theta^j{}_a$. Thus $P_i$ is the classical mechanical momentum resolved along the selected frame, with \cref{frame_momentum_symmetric_from_canonical} specifying the quantum ordering adopted here.

The uncertainty relations studied below pair the coordinate positions $x^i$ with these selected frame-momentum components. Their modified commutator reflects the position dependence of the frame, while $\widehat p_i^{\rm can}$ remains canonically conjugate to the same coordinate position $x^i$.

\subsection{Initial symmetric frame momenta}
\label{subsec_initial_symmetric_frame_momenta}

The symmetric ordering introduced in \cref{frame_momentum_symmetric_from_canonical} can also be obtained directly by integration by parts. For a positive measure density $\mu(x)$ and a real frame $E_i{}^a$, the first-order expression
\begin{align}
	\widehat P_i^{(\mathcal A)}
	=
	-i\hbar\left[
	E_i{}^a\partial_a
	+\frac{1}{2\mu}\partial_a(\mu E_i{}^a)
	\right]
	+\mathcal A_i(x)~,
	\qquad
	\mathcal A_i(x)\in\mathbb R~,
	\label{general_symmetric_momentum_operator}
\end{align}
satisfies
\begin{align}
	\langle\psi,\widehat P_i^{(\mathcal A)}\phi\rangle_\mu
	-
	\langle\widehat P_i^{(\mathcal A)}\psi,\phi\rangle_\mu
	=
	-i\hbar\int_{\partial\Omega}d\Sigma_a\,
	\mu E_i{}^a\psi^*\phi~,
	\label{momentum_boundary_flux}
\end{align}
where the inner products are evaluated over $\Omega$ and the coefficients and wavefunctions are assumed sufficiently regular there. For $\psi,\phi\in\mathcal D_0$, the region $\Omega$ can be chosen to contain their supports while excluding the center, so the boundary contribution vanishes. Symmetry therefore fixes the divergence correction for the specified frame and measure, while allowing an additional real multiplication term.

We require the additional multiplication term to transform as a vector under the rotations and reflections specified in \cref{frame_Od_equivariance}, which restricts it to $\mathcal A_i(x)=\alpha(r)n_i$ with real $\alpha(r)$. For the direct radial frame, let $\chi(r)$ be real and smooth for $r>0$, so that multiplication by $U_\chi=e^{i\chi(r)/\hbar}$ preserves $\mathcal D_0$ and gives
\begin{align}
	U_\chi\widehat P_i^{(\mathcal A)}U_\chi^{-1}
	&=
	\widehat P_i^{(\mathcal A')}~,
	&
	\mathcal A_i'
	&=
	\left[\alpha(r)-J(r)\chi'(r)\right]n_i~,
	\label{radial_phase_momentum_shift}
\end{align}
where $\chi$ has dimensions of action. A sufficiently regular radial term can therefore be removed locally by choosing $\chi'=\alpha/J$, while a global equivalence requires the phase transformation to map the chosen operator domains appropriately. Consistent with the symmetric ordering in the preceding subsection, we adopt the representative
\begin{align}
	\mathcal A_i=0~.
	\label{minimal_momentum_choice}
\end{align}

For the frame in \cref{direct_radial_frame} and the density $\mu=w_k=J^{-1}$, the two contributions to the divergence correction are
\begin{align}
	\partial_aE_i{}^a
	&=
	kd\,n_i~,
	&
	E_i{}^a\partial_a\log w_k
	&=
	-k\,n_i~,
	\qquad r>0~.
	\label{direct_radial_divergence_components}
\end{align}
Consequently,
\begin{align}
	\frac{1}{2w_k}\partial_a(w_kE_i{}^a)
	&=
	\frac12\left[
	\partial_aE_i{}^a
	+
	E_i{}^a\partial_a\log w_k
	\right]
	\nonumber\\
	&=
	\frac{k(d-1)}2n_i~,
	\label{direct_radial_connection_term}
\end{align}
and the initial frame-momentum operators become
\begin{align}
	P_{i,0}
	=
	-i\hbar\left[
	\partial_i+kx_i\partial_r+\frac{k(d-1)}2n_i
	\right]~,
	\label{direct_radial_symmetric_momentum}
\end{align}
on the common domain introduced in \cref{punctured_common_test_domain},
\begin{align}
	\mathcal D_0
	=
	C_c^\infty(\mathbb R^d\setminus\{0\})
	\subset\mathcal H_k~.
	\label{initial_momentum_domain}
\end{align}
The coefficients are smooth on the support of every wavefunction in $\mathcal D_0$, and \cref{momentum_boundary_flux} establishes symmetry on this domain. The corresponding self-adjoint generators require a separate construction, which is given in the following section.

Since the multiplication term in \cref{direct_radial_symmetric_momentum} commutes with position, the mixed commutator on $\mathcal D_0$ is determined by the frame,
\begin{align}
	[x^j,P_{i,0}]
	=
	i\hbar\left(\delta_i{}^j+kr\,n_i n^j\right)
	=
	i\hbar\left(\delta_i{}^j+k\frac{x_i x^j}{r}\right)~.
	\label{direct_radial_mixed_commutator}
\end{align}
This expression relates the coordinate positions to the selected frame-momentum components and provides the starting point for the uncertainty analysis below.

The ordinary differential expressions are recovered locally as $k\to0^+$. On every compact region away from the center, the measure density and operator coefficients have the limits
\begin{align}
	J(r)&\longrightarrow1~,
	&
	d\mu_k&\longrightarrow d^dx~,
	&
	P_{i,0}\psi&\longrightarrow-i\hbar\partial_i\psi~,
	\qquad \psi\in\mathcal D_0~.
	\label{direct_radial_undeformed_limit}
\end{align}
For a fixed $\psi\in\mathcal D_0$, the last limit also holds in the ordinary $L^2$ norm because its support is compact. Convergence of the self-adjoint operators requires a separate analysis using the unitary identification of the $k$-dependent Hilbert spaces in \cref{density_unitary_map_canonical}, and does not follow from these local limits alone.

\section{Momentum generators, radial uncertainty and dynamics}
\label{sec:momenta_uncertainty}

Throughout this section we consider $k>0$ and integer $d\geq2$, with $k\to0^+$ defining the undeformed limit. Starting from the measure and symmetric frame momenta introduced in the previous section, we now construct the self-adjoint frame momenta, determine their algebra and specify the frame Hamiltonian with its radial boundary conditions. We also derive the uncertainty bound and examine the radial volume scaling. Finally, we study the harmonic oscillator to determine the energy shifts and shell splittings induced by the deformation.

\subsection{Self-adjoint Cartesian frame-momentum generators}
\label{subsec_self_adjoint_momentum_generators}

We now construct self-adjoint realizations of the symmetric frame momenta introduced in \cref{subsec_initial_symmetric_frame_momenta} by identifying the finite displacements that they generate on the Hilbert space described by \cref{direct_radial_hilbert_space}. The derivative part of each momentum defines the vector field
\begin{align}
	X_i
	\equiv E_i{}^a\partial_a
	=
	\partial_i+kx_i\partial_r~,
	\label{direct_radial_vector_fields}
\end{align}
whose coefficients extend continuously to the center through $x_i x^a/r=0$ at $r=0$. This displacement combines the ordinary Cartesian direction with an additional radial contribution determined by the selected frame. Although the coefficients are not differentiable at the center, their variation between any two points is bounded by a fixed constant times the distance between those points. Indeed, the continuous extension $F_i(x)=x_i x/|x|$, with $F_i(0)=0$, satisfies $|F_i(x)-F_i(y)|\leq3|x-y|$, while the coefficient vector of $X_i$ obeys
\begin{align}
	|X_i(x)|
	\leq1+k|x_i|
	\leq1+kr~.
	\label{vector_field_linear_growth}
\end{align}
These properties ensure that the displacement trajectory through every point exists uniquely for all real values of its parameter,
\begin{align}
	\frac{d}{dt}\Phi_i^t(x)
	&=
	X_i\bigl(\Phi_i^t(x)\bigr)~,
	&
	\Phi_i^0(x)&=x~.
	\label{momentum_flow_equation}
\end{align}
Here $t$ has dimensions of length and parametrizes a displacement, rather than physical time. The linear growth bound excludes escape to infinity at finite $t$, and uniqueness gives $\Phi_i^{t+s}=\Phi_i^t\circ\Phi_i^s$ with inverse $\Phi_i^{-t}$.

A displacement changes both the coordinate volume and the radial weight $w_k$ in the inner product, so its action on a wavefunction must include the corresponding amplitude factor. We define
\begin{align}
	\rho_i(t,x)
	&=
	\frac{
		w_k\bigl(\Phi_i^{-t}(x)\bigr)
		\left|\det D\Phi_i^{-t}(x)\right|
	}{
		w_k(x)
	}~,
	\label{flow_density_factor}\\
	(U_i(t)\psi)(x)
	&=
	\rho_i(t,x)^{1/2}
	\psi\bigl(\Phi_i^{-t}(x)\bigr)~.
	\label{weighted_flow_unitary}
\end{align}
For each fixed $t$, the flow and its inverse are Lipschitz, so the Jacobian exists almost everywhere and the change-of-variables formula establishes
\begin{align}
	\|U_i(t)\psi\|_k
	&=
	\|\psi\|_k~,
	&
	U_i(t+s)&=U_i(t)U_i(s)~.
	\label{weighted_flow_group_law}
\end{align}
The square-root factor in \cref{weighted_flow_unitary} therefore preserves probability as the wavefunction is displaced through regions with different volume weight. For states in $\mathcal D_0$, smoothness of the flow near their compact support gives $U_i(t)\psi\to\psi$ in norm as $t\to0$, and unitarity extends this continuity to the full Hilbert space.

Stone's theorem then determines a unique self-adjoint generator for this unitary group \cite{Stone:1932,ReedSimonI},
\begin{align}
	U_i(t)
	=
	\exp\left(-\frac{it}{\hbar}\widehat P_i\right)~,
	\label{Stone_generator_definition}
\end{align}
with domain
\begin{align}
	\mathcal D(\widehat P_i)
	=
	\left\{
	\psi\in\mathcal H_k
	\;\middle|\;
	\lim_{t\to0}
	\frac{U_i(t)\psi-\psi}{t}
	\text{ exists in }\mathcal H_k
	\right\}~.
	\label{Stone_generator_domain}
\end{align}
The globally Lipschitz extension fixes the flow through the center, so this construction introduces no independent boundary condition there. The uniqueness asserted here concerns the generator of the specified unitary group and does not require a separate claim that the initial symmetric operator is essentially self-adjoint.

Differentiating \cref{weighted_flow_unitary} on $\mathcal D_0$ and using the measure correction in \cref{direct_radial_connection_term}, we obtain
\begin{align}
	\left.\frac{d}{dt}U_i(t)\psi\right|_{t=0}
	&=
	-\left[
	X_i+
	\frac{1}{2w_k}\partial_a(w_kE_i{}^a)
	\right]\psi~,
	\label{weighted_flow_infinitesimal_generator}\\
	\widehat P_i\psi
	&=
	-i\hbar\left[
	X_i+\frac{k(d-1)}{2}n_i
	\right]\psi
	=
	P_{i,0}\psi~,
	\qquad \psi\in\mathcal D_0~.
	\label{self_adjoint_momentum_local_action}
\end{align}
Thus the finite displacement reproduces the symmetric momentum expression in \cref{direct_radial_symmetric_momentum}, while \cref{Stone_generator_domain} supplies its self-adjoint realization. The local commutators below will be evaluated on $\mathcal D_0$, where the differential expressions and their products are well defined.

Rotations preserve the radial measure and act unitarily as
\begin{align}
	(V(R)\psi)(x)
	&=
	\psi(R^{-1}x)~,
	\qquad R\in SO(d)~,
	\label{unitary_rotation_representation}\\
	\widehat L_{ij}
	&=
	-i\hbar(x_i\partial_j-x_j\partial_i)~,
	\qquad \text{on }\mathcal D_0~.
	\label{angular_momentum_definition}
\end{align}
To express covariance without imposing additional assumptions on sums of unbounded operators, let $\widehat P(a)$ denote the self-adjoint generator constructed from $X_a=a^iX_i$ for a constant dimensionless vector $a$. The same flow construction applies to $X_a$, and the frame covariance in \cref{frame_Od_equivariance} gives
\begin{align}
	V(R)\widehat P(a)V(R)^{-1}
	&=
	\widehat P(Ra)~,
	\label{closed_generator_rotation_covariance}\\
	V(R)\mathcal D\bigl(\widehat P(a)\bigr)
	&=
	\mathcal D\bigl(\widehat P(Ra)\bigr)~.
	\label{closed_generator_rotation_domain}
\end{align}
Since $\widehat P(a)$ acts as $a^i\widehat P_i$ on $\mathcal D_0$, the component relation is
\begin{align}
	V(R)\widehat P_iV(R)^{-1}
	=
	R^j{}_i\widehat P_j~,
	\qquad \text{on }\mathcal D_0~.
	\label{local_momentum_vector_transformation}
\end{align}
The resulting observables therefore respect the rotational covariance of the construction, although their unitary displacements need not preserve a chosen Hamiltonian and do not imply translation symmetry of the radial background.

\subsection{Algebra of frame momenta}
\label{subsec_momentum_algebra}

The local actions of $x^j$, $\widehat P_i$ and $\widehat L_{ij}$ preserve $\mathcal D_0$, so both orders of every operator product considered below are defined on this common domain. All commutators in this subsection are evaluated on the common domain $\mathcal D_0$, where the local action in \cref{self_adjoint_momentum_local_action} gives
\begin{align}
	[x^j,\widehat P_i]
	=
	i\hbar E_i{}^j~,
	\label{mixed_commutator_section4}
\end{align}
in agreement with \cref{direct_radial_mixed_commutator}. The frame vector fields in \cref{direct_radial_vector_fields} satisfy
\begin{align}
	[X_i,X_j]
	=
	-\frac{k}{r}
	\left(x_i\partial_j-x_j\partial_i\right)~.
	\label{vector_field_anholonomy}
\end{align}
Thus the commutator of two infinitesimal frame displacements acts tangentially to the spheres centred at the origin, with a rotation coefficient that depends on the radius. The multiplication terms in \cref{self_adjoint_momentum_local_action} produce no additional contribution because
\begin{align}
	X_i n_j
	=
	\frac{\delta_{ij}-n_i n_j}{r}
	=
	X_j n_i~.
	\label{frame_derivative_radial_unit_vector}
\end{align}
Using the angular momentum convention in \cref{angular_momentum_definition}, we therefore obtain
\begin{align}
	[\widehat P_i,\widehat P_j]
	=
	i\hbar\frac{k}{r}\widehat L_{ij}~.
	\label{momentum_momentum_commutator}
\end{align}
Since $r^{-1}$ commutes with $\widehat L_{ij}$ on $\mathcal D_0$, its placement in this expression introduces no ordering ambiguity. The coefficient depends on position, so the momentum and angular momentum generators do not close among themselves with constant coefficients.

The rotational transformation law established in \cref{local_momentum_vector_transformation} gives
\begin{align}
	[\widehat L_{ij},\widehat P_m]
	=
	i\hbar
	\left(
	\delta_{im}\widehat P_j
	-
	\delta_{jm}\widehat P_i
	\right)~.
	\label{rotation_momentum_commutator}
\end{align}
The noncommutativity of the frame momenta is already present classically. Indeed, the canonical brackets in \cref{canonical_phase_space} and the definition in \cref{physical_momentum} imply
\begin{align}
	\{P_i,P_j\}
	=
	\frac{k}{r}L_{ij}~,
	\qquad
	L_{ij}=x_i p_j-x_j p_i~.
	\label{classical_frame_momentum_bracket}
\end{align}
The quantum relation \cref{momentum_momentum_commutator} therefore reflects the position dependence of the chosen frame, rather than an additional term generated by the Hilbert-space measure.

Although each self-adjoint component $\widehat P_i$ admits a spectral representation, the full family cannot be diagonalized simultaneously. To see this, suppose that a single unitary transformation $\mathcal F$ represented every component as multiplication by a real momentum coordinate,
\begin{align}
	\mathcal F\widehat P_i\mathcal F^{-1}
	=
	M_{p_i}~,
	\qquad
	(M_{p_i}\widetilde\psi)(p)
	=
	p_i\widetilde\psi(p)~.
	\label{hypothetical_joint_momentum_representation}
\end{align}
Such multiplication operators have commuting spectral projections, and their unitary equivalents would therefore strongly commute. In particular, the two orders of their products would agree on $\mathcal D_0$, where both are defined. This contradicts \cref{momentum_momentum_commutator}, whose right-hand side is nonzero on test functions with suitable angular dependence. Hence, for $k\neq0$ and $d\geq2$, there is no joint Cartesian spectral representation in which all frame momenta act as commuting multiplication variables.

This obstruction concerns the frame momenta $\widehat P_i$. The canonical momenta in \cref{canonical_momentum_domain} retain their joint Fourier representation, obtained by applying the density map $W$ in \cref{density_unitary_map_canonical} followed by the ordinary Cartesian Fourier transform. Likewise, individual frame components remain accessible through their own spectral representations. For a self-adjoint Hamiltonian whose operator domain and action respect rotations, the energy can instead be resolved together with a commuting set of angular observables. In three dimensions these may be chosen as $\widehat{\mathbf L}^{\,2}$ and $\widehat L_{12}$, while in general dimension the corresponding angular labels are supplied by hyperspherical harmonics. Such an energy and angular momentum representation is compatible with \cref{momentum_momentum_commutator}, since it does not require simultaneous diagonalization of the Cartesian frame momenta.

\subsection{Radial phase modes and the half-line obstruction}
\label{subsec_radial_characters_obstruction}

The additive coordinate introduced in \cref{radial_additive_coordinate} suggests the radial phase modes
\begin{align}
	K_p(r)
	=
	\exp\left(\frac{ip}{\hbar}\tau_k(r)\right)
	=
	(1+kr)^{ip/(\hbar k)}~,
	\qquad p\in\mathbb R~.
	\label{radial_phase_kernel}
\end{align}
By \cref{radial_additive_law,radial_Borges_derivative}, these modes obey 
\begin{align}
	K_p(r\oplus_k a)
	&=
	K_p(r)K_p(a)~,
	\label{radial_phase_composition}\\
	-i\hbar D_k^{(r)}K_p(r)
	&=
	pK_p(r)~.
	\label{radial_phase_eigenvalue}
\end{align}
Although these relations reproduce the local plane-wave behaviour in $\tau_k$, identifying $p$ with the spectrum of a radial momentum requires accounting for both the radial measure and the endpoint at $r=0$.

On $\mathcal D_0$, the symmetric radial projection of the frame momenta is
\begin{align}
	\Pi_{r,0}
	&=
	\frac12
	\left(n^iP_{i,0}+P_{i,0}n^i\right)
	\nonumber\\
	&=
	-i\hbar J(r)
	\left(
	\partial_r+\frac{d-1}{2r}
	\right)~.
	\label{symmetric_radial_momentum}
\end{align}
The additional term follows from the symmetric projection and is consistent with the radial probability measure. Indeed, the Hilbert space in \cref{direct_radial_hilbert_space} decomposes as
\begin{align}
	\mathcal H_k
	\simeq
	L^2\left(
	(0,\infty),\frac{r^{d-1}dr}{J(r)}
	\right)
	\otimes
	L^2(S^{d-1},d\Omega)~,
	\label{full_radial_angular_decomposition}
\end{align}
so the transformation
\begin{align}
	(U_r\psi)(\tau_k,\Omega)
	=
	r(\tau_k)^{(d-1)/2}
	\psi(r(\tau_k),\Omega)
	\label{radial_unitary_flattening_map}
\end{align}
is unitary from $\mathcal H_k$ to $L^2((0,\infty),d\tau_k)\otimes L^2(S^{d-1},d\Omega)$, where $r(\tau_k)$ is the inverse radial coordinate in \cref{tau_metric_warped_form}. Its action on the initial radial operator gives
\begin{align}
	U_r\Pi_{r,0}U_r^{-1}
	=
	-i\hbar\partial_{\tau_k}\otimes\mathbf1
	\label{radial_momentum_half_line_derivative}
\end{align}
on the transformed test domain.

The physical radial range remains a half-line, as established in \cref{radial_half_line_map}. The operator $-i\hbar\partial_{\tau_k}$ initially defined on $C_c^\infty(0,\infty)$ has no self-adjoint extension on $L^2(0,\infty)$ \cite{Bonneau:2001,ReedSimonII}.\footnote{For $T_0=-i\hbar\partial_{\tau_k}$, the equations $(T_0^*\mp i\hbar\lambda)f_\pm=0$, with $\lambda>0$ of dimension $L^{-1}$, have solutions $f_\pm=C_\pm e^{\mp\lambda\tau_k}$. Only $f_+$ is square integrable, giving deficiency indices $(1,0)$ for each fixed angular mode and $(\infty,0)$ on the full radial-angular Hilbert space. Their inequality excludes a self-adjoint extension.} Consequently, no choice of endpoint condition turns $\Pi_{r,0}$ into a self-adjoint radial momentum with this first-order action on $\mathcal H_k$. This result reflects the restriction of radial displacements to the positive half-line and persists in the undeformed limit \cite{Paz:2002}. It is therefore compatible with the self-adjoint Cartesian frame generators constructed in \cref{subsec_self_adjoint_momentum_generators}.

The formal eigenfunctions of the local radial expression include the amplitude required by \cref{radial_unitary_flattening_map},
\begin{align}
	\chi_p(r)
	=
	r^{-(d-1)/2}K_p(r)~,
	\qquad
	\Pi_{r,0}\chi_p=p\chi_p
	\quad\text{as a differential equation for }r>0~.
	\label{formal_radial_momentum_eigenfunctions}
\end{align}
They do not constitute the spectral basis of a self-adjoint radial momentum, although Fourier integrals involving these phases remain available as representations of radial wavefunctions. In particular, Fourier transforming the zero extension of $U_r\psi$ preserves its norm but maps onto only a proper subspace of the full Fourier-space Hilbert space.

For the dynamics considered below, the radial spectral representation will instead be determined by the self-adjoint Hamiltonian and its endpoint conditions. The transformation \cref{radial_unitary_flattening_map} remains useful for this purpose because it removes the radial weight and simplifies the radial derivative, without assuming the existence of a self-adjoint first-order radial momentum.

\subsection{Frame Hamiltonian and radial boundary conditions}
\label{subsec_global_frame_hamiltonian}

The self-adjoint frame momenta constructed in \cref{subsec_self_adjoint_momentum_generators} allow a kinetic energy to be defined, although the kinematics does not uniquely select its quantum prescription. We choose the Hamiltonian associated with the sum of the squared frame-momentum components through
\begin{align}
	\mathfrak q_P[\psi]
	&=
	\frac{1}{2m}
	\sum_{i=1}^d
	\|\widehat P_i\psi\|_k^2~,
	&
	\mathcal D(\mathfrak q_P)
	&=
	\bigcap_{i=1}^d
	\mathcal D(\widehat P_i)~.
	\label{initial_frame_hamiltonian_form}
\end{align}
This domain contains $\mathcal D_0$ and is therefore dense in $\mathcal H_k$. Since each $\widehat P_i$ is closed, the form is closed and non-negative, and hence determines a unique non-negative self-adjoint Hamiltonian $\widehat H_P$ \cite{Kato:1995}. As shown in \cref{app_weak_domain_form_core}, every state in $\mathcal D(\mathfrak q_P)$ can be approximated by states in $\mathcal D_0$ such that both the norm and the kinetic energy of their difference tend to zero. More precisely, with
\begin{align}
	\|\psi\|_{P,1}^2
	=
	\|\psi\|_k^2
	+
	\frac{\mathfrak q_P[\psi]}{E_*}~,
	\qquad E_*>0~,
	\label{frame_form_norm_equivalent}
\end{align}
where $E_*$ is an arbitrary reference energy, the approximation statement is
\begin{align}
	\overline{\mathcal D_0}^{\,\|\cdot\|_{P,1}}
	=
	\mathcal D(\mathfrak q_P)~.
	\label{D0_form_core_statement}
\end{align}
Thus closing the local form
\begin{align}
	\mathfrak q_{P,0}[\psi]
	=
	\frac{1}{2m}
	\sum_{i=1}^d
	\|P_{i,0}\psi\|_k^2~,
	\qquad \psi\in\mathcal D_0~,
	\label{local_frame_hamiltonian_form}
\end{align}
recovers \cref{initial_frame_hamiltonian_form}. This identifies $\widehat H_P$ with the Friedrichs realization selected by the local kinetic-energy form, without imposing an independent interaction at the center.

To obtain its differential action, define
\begin{align}
	\widehat{\mathbf P}^{\,2}
	&=
	\sum_{i=1}^d\widehat P_i^2~,
	&
	\widehat L^2
	&=
	\frac12\sum_{i,j=1}^d\widehat L_{ij}^2
	\label{P2_L2_definitions}
\end{align}
on $\mathcal D_0$. Since the momentum operators preserve this domain, the Hamiltonian acts there as
\begin{align}
	\widehat H_P\psi
	=
	\frac{1}{2m}
	\widehat{\mathbf P}^{\,2}\psi~,
	\qquad \psi\in\mathcal D_0~.
	\label{frame_hamiltonian_local_action}
\end{align}
Using the symmetric radial expression in \cref{symmetric_radial_momentum}, direct evaluation gives
\begin{align}
	\widehat{\mathbf P}^{\,2}
	=
	\Pi_{r,0}^2
	+
	\frac{\widehat L^2}{r^2}
	+
	\frac{\hbar^2(d-1)(d-3)}{4r^2}
	-
	\frac{\hbar^2k(d-1)}{2r}~.
	\label{radial_decomposition_p_squared}
\end{align}
Here $\Pi_{r,0}^2$ denotes the local second-order expression on $\mathcal D_0$, without assigning a self-adjoint first-order radial momentum.

The relation to geometric quantization follows from the classical identity
\begin{align}
	H_{\rm cl}
	=
	\frac{\delta^{ij}P_iP_j}{2m}
	=
	\frac{g^{ab}p_ap_b}{2m}~.
	\label{classical_frame_metric_kinetic_equivalence}
\end{align}
For the metric in \cref{direct_radial_line_element}, the scalar Laplacian is
\begin{align}
	\Delta_g
	=
	\frac{J}{r^{d-1}}
	\partial_r
	\left(r^{d-1}J\partial_r\right)
	+
	\frac{1}{r^2}\Delta_{S^{d-1}}~,
	\label{frame_metric_scalar_laplacian}
\end{align}
and the frame kinetic expression satisfies
\begin{align}
	\widehat{\mathbf P}^{\,2}
	=
	-\hbar^2\Delta_g
	-
	\hbar^2
	\left[
	\frac{k(d-1)^2}{2r}
	+
	\frac{k^2(d-1)^2}{4}
	\right]
	\qquad\text{on }\mathcal D_0~.
	\label{p_squared_laplace_beltrami_relation}
\end{align}

For comparison, the geometric Hamiltonian $\widehat H_g^F$ is defined by closing
\begin{align}
	\mathfrak q_{g,0}[\psi]
	=
	\frac{\hbar^2}{2m}
	\int_{\mathbb R^d}d\mu_k\,
	g^{ab}\partial_a\psi^*\partial_b\psi~,
	\qquad \psi\in\mathcal D_0~.
	\label{initial_geometric_hamiltonian_form}
\end{align}
Its local action is $-\hbar^2\Delta_g/(2m)$, so \cref{p_squared_laplace_beltrami_relation} gives
\begin{align}
	\widehat H_P^{\rm loc}
	=
	\widehat H_g^{\rm loc}
	-
	\frac{\hbar^2}{2m}
	\left[
	\frac{k(d-1)^2}{2r}
	+
	\frac{k^2(d-1)^2}{4}
	\right]~.
	\label{frame_vs_laplace_hamiltonian_local}
\end{align}
This comparison concerns the local differential expressions, while their self-adjoint realizations are specified by the respective forms. The two prescriptions share the classical kinetic energy in \cref{classical_frame_metric_kinetic_equivalence} but differ by a quantum potential that can affect their spectra. The attractive sign in \cref{frame_vs_laplace_hamiltonian_local} does not compromise the non-negativity of $\widehat H_P$, which follows from \cref{initial_frame_hamiltonian_form}. The scalar $q$-calculus does not select between these prescriptions, and all subsequent predictions based on $\widehat H_P$ refer to the frame prescription chosen here.

The frame form is rotationally invariant on $\mathcal D_0$, and the approximation result \cref{D0_form_core_statement} extends this invariance to its full domain,
\begin{align}
	V(R)\mathcal D(\mathfrak q_P)
	&=
	\mathcal D(\mathfrak q_P)~,
	&
	\mathfrak q_P[V(R)\psi]
	&=
	\mathfrak q_P[\psi]~.
	\label{frame_form_rotation_invariance}
\end{align}
Consequently, $\widehat H_P$ commutes with the unitary rotations, including the corresponding transformation of its operator domain, and the spectral problem separates into angular momentum sectors. We write
\begin{align}
	\psi(r,\Omega)
	&=
	\sum_{\ell,\alpha}
	\psi_{\ell\alpha}(r)Y_{\ell\alpha}(\Omega)~,
	\label{partial_wave_decomposition}\\
	\widehat L^2Y_{\ell\alpha}
	&=
	\hbar^2\ell(\ell+d-2)Y_{\ell\alpha}~,
	\label{angular_casimir_eigenvalues}
\end{align}
where $\alpha$ labels the angular degeneracy.

Applying the unitary map in \cref{radial_unitary_flattening_map}, with $u_{\ell\alpha}(\tau_k) =r(\tau_k)^{(d-1)/2}\psi_{\ell\alpha}(r(\tau_k))$, gives the radial expression
\begin{align}
	\widehat H_{P,\ell}^{\rm loc}
	=
	\frac{\hbar^2}{2m}
	\left[
	-\frac{d^2}{d\tau_k^2}
	+
	\frac{\nu_\ell^2-\frac14}{r(\tau_k)^2}
	-
	\frac{k(d-1)}{2r(\tau_k)}
	\right]~,
	\qquad
	\nu_\ell=\ell+\frac{d-2}{2}~.
	\label{partial_wave_frame_hamiltonian}
\end{align}
The self-adjoint radial Hamiltonian is obtained by closing the corresponding form initially defined on $C_c^\infty(0,\infty)$, as inherited from the global construction.

Near the center, $r(\tau_k)=\tau_k+O(k\tau_k^2)$, so the leading
singularity is $(\nu_\ell^2-\frac14)/\tau_k^2$. The remaining terms are at most of order $1/\tau_k$ and do not change which local solutions are square integrable at the endpoint. For $0<\nu_\ell<1$, two independent solutions of the radial eigenvalue equation have leading behaviours
\begin{align}
	u_+(\tau_k)
	&\sim \tau_k^{1/2+\nu_\ell}~,
	&
	u_-(\tau_k)
	&\sim \tau_k^{1/2-\nu_\ell}~,
	\label{inverse_square_local_behaviour_general}
\end{align}
and the Friedrichs prescription sets the coefficient of $u_-$ to zero. At $\nu_\ell=0$, the corresponding behaviours are
\begin{align}
	u_+(\tau_k)
	&\sim \tau_k^{1/2}~,
	&
	u_-(\tau_k)
	&\sim
	\tau_k^{1/2}\log(\tau_k/\ell_*)~,
	\label{critical_inverse_square_behaviour}
\end{align}
where $\ell_*>0$ is an arbitrary reference length, and the logarithmic branch is excluded. These asymptotics describe local eigenfunctions, while the full form domain is defined by closure. In particular, at the critical inverse-square coupling, the derivative and inverse-square contributions should not be required to be separately finite.

For integer $d\geq2$, both local solutions are square integrable only in the $s$-wave sectors of $d=2$ and $d=3$. The chosen form selects
\begin{align}
	u(\tau_k)
	&\sim C\,\tau_k^{1/2}~,
	&& d=2,\quad\ell=0~,
	\nonumber\\
	u(\tau_k)
	&\sim C\,\tau_k~,
	&& d=3,\quad\ell=0~,
	\label{frame_friedrichs_s_wave_conditions}
\end{align}
with the logarithmic branch excluded in the first case and the branch approaching a nonzero constant excluded in the second.
For $\nu_\ell\geq1$, only one local solution is square integrable, so no independent boundary condition is available at the center for the same radial expression \cite{ReedSimonII,Schmudgen:2012}.

For the free frame Hamiltonian considered here, the radial potential in \cref{partial_wave_frame_hamiltonian} vanishes exponentially as $\tau_k\to\infty$, since $r(\tau_k)=(e^{k\tau_k}-1)/k$. Infinity therefore requires no additional self-adjoint boundary condition. The global frame form thus fixes the radial Hamiltonian in every angular sector. Alternative center conditions are possible for the same local expression only in the two exceptional $s$-wave sectors described above, and would define different Hamiltonians rather than an ambiguity remaining in $\widehat H_P$. Whether such alternatives describe a microscopic core requires additional physical input beyond the scalar radial deformation.

\subsection{Position-dependent uncertainty bounds}
\label{subsec_position_momentum_uncertainty}

We now derive the uncertainty relation for the Cartesian coordinate positions $x^i$ and the self-adjoint frame momenta $\widehat P_i$. Their distinction from the canonical pairs was established in \cref{subsec_canonical_local_frame_momenta}, and the resulting bound refers to the selected radial frame and its physical center. Throughout this subsection, $\psi\in\mathcal D_0$ is normalized, and all expectation values and variances are evaluated in this state. This domain ensures that the operator products entering the commutators are well defined. For a fixed Cartesian component, Robertson's inequality \cite{Robertson:1929} and \cref{mixed_commutator_section4} give
\begin{align}
	\Delta x^i\,\Delta P_i
	&\geq
	\frac12
	\left|
	\left\langle
	[x^i,\widehat P_i]
	\right\rangle
	\right|
	\nonumber\\
	&=
	\frac{\hbar}{2}
	\left[
	1+k
	\left\langle
	\frac{(x^i)^2}{r}
	\right\rangle
	\right]~,
	\qquad \text{no sum over }i~.
	\label{directional_radial_gup}
\end{align}
The absolute value has been removed because $k>0$ and $(x^i)^2/r\geq0$. The directional dependence enters through $(x^i)^2/r=r n_i^2$, which weights the radial distribution by its projection along the chosen Cartesian axis. If the position probability distribution is rotationally symmetric, then
\begin{align}
	\left\langle
	\frac{(x^i)^2}{r}
	\right\rangle
	=
	\frac{\langle r\rangle}{d}~,
	\label{isotropic_state_radial_expectation}
\end{align}
since the contributions from all directions are equal and their sum is $\langle r\rangle$. This condition concerns the probability distribution and does not require the phase of the wavefunction to be rotationally invariant. In this case each component obeys
\begin{align}
	\Delta x^i\,\Delta P_i
	\geq
	\frac{\hbar}{2}
	\left(
	1+\frac{k}{d}\langle r\rangle
	\right)~.
	\label{rotationally_symmetric_directional_gup}
\end{align}

A rotationally invariant combination of the component uncertainties can also be bounded without imposing any symmetry on the state where we define
\begin{align}
	\Delta x_{\rm av}
	&=
	\left[
	\frac1d\sum_{i=1}^d(\Delta x^i)^2
	\right]^{1/2}~,
	&
	\Delta P_{\rm av}
	&=
	\left[
	\frac1d\sum_{i=1}^d(\Delta P_i)^2
	\right]^{1/2}~.
	\label{averaged_uncertainties_definition}
\end{align}
These quantities are invariant under global rotations of the coordinate and frame components. In particular,
\begin{align}
	\Delta x_{\rm av}^{\,2}
	=
	\frac{
		\langle r^2\rangle
		-
		|\langle\mathbf x\rangle|^2
	}{d}~,
	\label{averaged_position_variance}
\end{align}
which distinguishes the spatial spread of the state from its mean distance to the selected center. Summing \cref{directional_radial_gup} over $i$ and applying the Cauchy-Schwarz inequality gives
\begin{align}
	d\,\Delta x_{\rm av}\Delta P_{\rm av}
	&\geq
	\sum_{i=1}^d\Delta x^i\,\Delta P_i
	\nonumber\\
	&\geq
	\frac{\hbar}{2}
	\left(d+k\langle r\rangle\right)~,
	\label{summed_directional_gup}
\end{align}
and consequently,
\begin{align}
		\Delta x_{\rm av}\Delta P_{\rm av}
		\geq
		\frac{\hbar}{2}
		\left(
		1+\frac{k}{d}\langle r\rangle
		\right)~.
	\label{radial_scale_dependent_gup}
\end{align}
This result holds for arbitrary normalized states in $\mathcal D_0$, including states with angular dependence. Since its derivation combines several inequalities, it does not establish that the lower bound is optimal or that it can be saturated.

For a state supported in a shell $R_0-\sigma<r<R_0+\sigma$, with $0<\sigma<R_0$, one has $R_0-\sigma\leq\langle r\rangle\leq R_0+\sigma$. The expression in \cref{radial_scale_dependent_gup} therefore implies
\begin{align}
	\Delta x_{\rm av}\Delta P_{\rm av}
	\geq
	\frac{\hbar}{2}
	\left[
	1+\frac{k}{d}(R_0-\sigma)
	\right]~.
	\label{shell_robertson_bound}
\end{align}
The shell support controls the mean radius without fixing the angular distribution or the actual uncertainty product. The dimensionless correction is governed by $k\langle r\rangle$, so it is small when the state occupies radial scales much smaller than $k^{-1}$.

The physical interpretation depends on the observables entering the bound. The expression in \cref{radial_scale_dependent_gup} relates coordinate-position uncertainties to momentum components in the preferred local frame, and therefore does not establish a universal modification of a local canonical uncertainty relation. Its radial dependence refers to the center specified by the model, rather than to an arbitrary choice of origin in homogeneous space. Position-dependent uncertainty relations also occur in earlier deformed-momentum constructions. The nonadditive displacement model of \cite{CostaFilho:2011} reports an uncertainty correction proportional to the mean position. Hermitian realizations of the corresponding deformed momentum were subsequently developed in \cite{Mazharimousavi:2012,DaCosta:2014}. In the scalar radial framework of \cite{Quesne:2004}, applying Robertson's inequality to the mixed commutator with a deformation factor linear in radius likewise produces a mean-radius correction. Thus the dependence on $\langle r\rangle$ alone does not distinguish the present model from all previous deformations. A comparison of physical predictions must also retain the specified position observables, momentum operators and Hamiltonian.

Finally, the model does not impose a universal minimum coordinate uncertainty. For any fixed $x_0\neq0$, normalized smooth wavefunctions can be supported in arbitrarily small balls about $x_0$ while remaining in $\mathcal D_0$. Their coordinate uncertainties tend to zero, with increasing momentum uncertainties consistent with \cref{radial_scale_dependent_gup}. The bound therefore constrains the uncertainty product at a given mean radial scale, but imposes no nonzero minimum coordinate width when no restriction on the kinetic energy is imposed.

\subsection{Radial volume scaling and hyperscaling interpretation}
\label{subsec_radial_volume_scaling}

The geometric measure in \cref{direct_radial_geometric_volume} determines the volume enclosed by a sphere of areal radius $R$ centred at the distinguished point,
\begin{align}
	V_k(R)
	=
	\int_{B_R}d\mu_k
	=
	\Omega_{d-1}
	\int_0^R\frac{r^{d-1}}{1+kr}\,dr~,
	\label{effective_spatial_capacity}
\end{align}
where $B_R=\{x\in\mathbb R^d\mid |x|<R\}$ denotes the ball of areal radius $R$ centred at the distinguished point and $\Omega_{d-1}$ is the area of the unit $(d-1)$-sphere. Since the angular part of the metric in \cref{direct_radial_line_element} is undeformed, the bounding sphere retains the area
\begin{align}
	A(R)=\Omega_{d-1}R^{d-1}~,
	\label{radial_boundary_area}
\end{align}
so the modification of the enclosed volume arises entirely from the radial distance element.

Introducing the dimensionless radius $K=kR$, we obtain
\begin{align}
	V_k(R)
	&=
	\Omega_{d-1}k^{-d}F_d(K)~,
	&
	F_d(K)
	&=
	\int_0^K\frac{t^{d-1}}{1+t}\,dt~.
	\label{capacity_Fd_definition}
\end{align}
For integer $d>1$, the integral evaluates to
\begin{align}
	F_d(K)
	=
	\sum_{m=0}^{d-2}
	(-1)^m\frac{K^{d-1-m}}{d-1-m}
	+
	(-1)^{d-1}\log(1+K)~.
	\label{capacity_Fd_integer}
\end{align}
At small areal radius, the expansion
\begin{align}
	V_k(R)
	=
	\frac{\Omega_{d-1}R^d}{d}
	\left[
	1-\frac{d}{d+1}kR
	+O\bigl((kR)^2\bigr)
	\right]~,
	\qquad kR\ll1~,
	\label{capacity_short_distance}
\end{align}
recovers the Euclidean volume at leading order. For $d>2$, the large-radius behaviour instead becomes
\begin{align}
	V_k(R)
	=
	\frac{\Omega_{d-1}R^{d-1}}{k(d-1)}
	\left[
	1+O\bigl((kR)^{-1}\bigr)
	\right]~,
	\qquad kR\gg1~.
	\label{capacity_large_distance_d_gt_2}
\end{align}
The exact result in two dimensions,
\begin{align}
	V_k(R)
	=
	\frac{\Omega_1}{k^2}
	\left[kR-\log(1+kR)\right]~,
	\label{capacity_large_distance_d_2}
\end{align}
has the same leading power with a logarithmic subleading correction. Consequently, for every integer $d\geq2$,
\begin{align}
	\frac{V_k(R)}{A(R)}
	\longrightarrow
	\frac{1}{k(d-1)}
	\qquad (kR\to\infty)~.
	\label{asymptotic_volume_area_ratio}
\end{align}
The enclosed volume is therefore asymptotically proportional to the boundary area, with the required length scale supplied by $k^{-1}$.

The crossover is described by the logarithmic growth exponent
\begin{align}
	\gamma_V(R)
	\equiv
	\frac{d\log V_k(R)}{d\log R}
	=
	\frac{K^d}{(1+K)F_d(K)}~,
	\label{capacity_dimension_exact}
\end{align}
whose limiting values are
\begin{align}
	\gamma_V(R)&\longrightarrow d~,
	&& kR\to0~,
	\label{capacity_dimension_short}\\
	\gamma_V(R)&\longrightarrow d-1~,
	&& kR\to\infty~.
	\label{capacity_dimension_large}
\end{align}
These limits describe growth with areal radius, whose relation to proper radial distance follows from \cref{radial_additive_coordinate,tau_metric_warped_form}. Defining
\begin{align}
	\mathcal V_k(\tau)
	\equiv
	V_k\bigl(R(\tau)\bigr)~,
	\qquad
	R(\tau)=\frac{e^{k\tau}-1}{k}~,
	\label{proper_radius_volume_definition}
\end{align}
we find
\begin{align}
	\mathcal V_k(\tau)
	\sim
	\frac{\Omega_{d-1}}{(d-1)k^d}
	e^{(d-1)k\tau}
	\qquad (k\tau\to\infty)~.
	\label{proper_radius_volume_asymptotic}
\end{align}
This exponential growth is consistent with the asymptotic negative curvature discussed in \cref{subsec_geometric_interpretation_measure}. The limit $\gamma_V\to d-1$ consequently does not establish a reduction of spatial dimension, although \cref{asymptotic_volume_area_ratio} remains a geometric relation between the volume and boundary area of centred balls.

The same volume determines the spatial factor in a semiclassical counting problem. Consider the frame Hamiltonian confined to $B_R$, with a Dirichlet condition at the outer boundary and the center prescription specified in \cref{subsec_global_frame_hamiltonian}. Using \cref{phase_space_measure_direct_radial,classical_frame_metric_kinetic_equivalence}, the leading semiclassical number of states below energy $E>0$ is
\begin{align}
	N_R^{\rm sc}(E)
	=
	\int_{\substack{x\in B_R\\
			\delta^{ij}P_iP_j\leq2mE}}
	d\Gamma_k=
	\frac{v_d(2mE)^{d/2}}{(2\pi\hbar)^d}
	V_k(R)~,
	\qquad
	v_d=\frac{\Omega_{d-1}}{d}~.
	\label{radial_volume_semiclassical_count}
\end{align}
This expression supplies the leading high-energy approximation at fixed $R$, with the deformation modifying the spatial factor while preserving the energy dependence $E^{d/2}$. The volume integral alone does not determine the complete spectrum or a global counting function for the noncompact problem.

\begin{figure}[t]
	\centering
	\includegraphics[width=\textwidth]{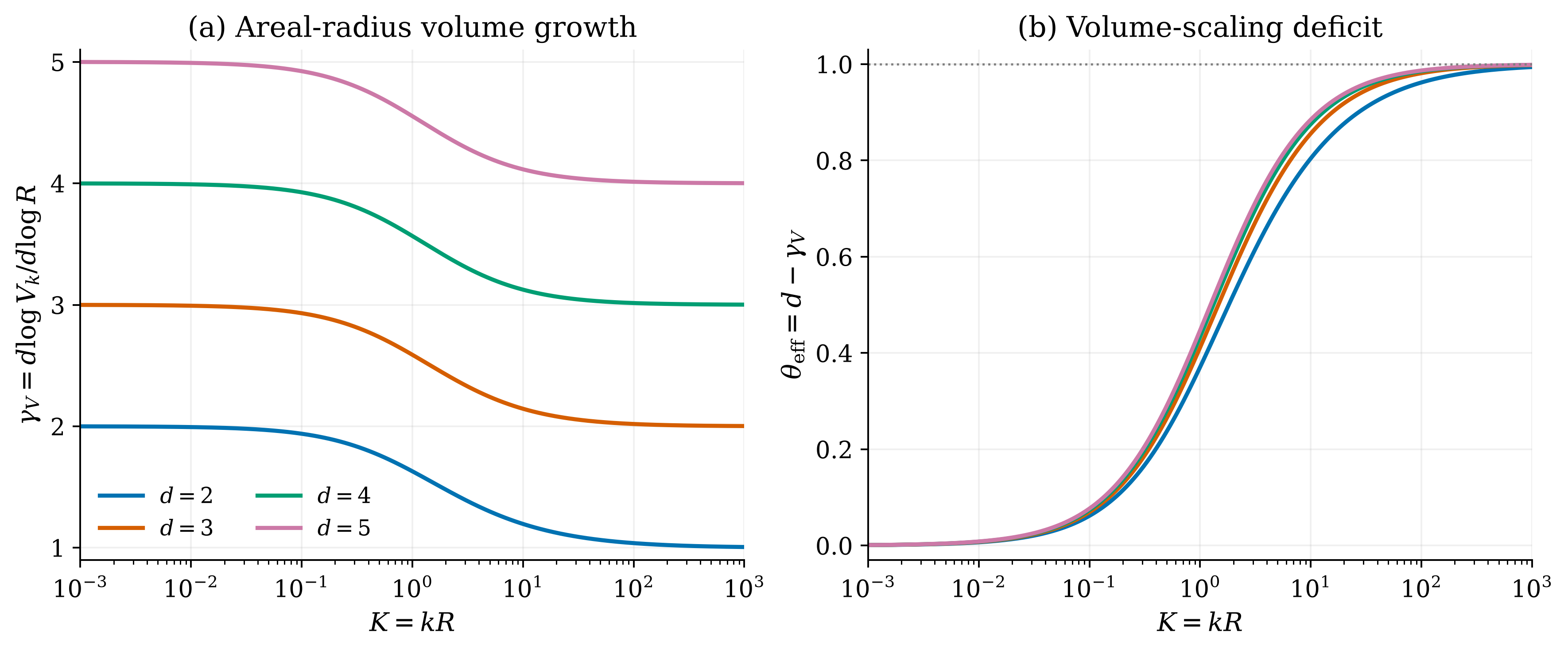}
	\caption{
		Radial volume scaling of the direct completion as a function of $K=kR$ for $d=2,3,4,5$, with the same colour assignment in both panels. The left panel shows the logarithmic growth exponent $\gamma_V$, which interpolates between $d$ at small areal radius and $d-1$ at large areal radius, while the right panel shows the corresponding quantity $\theta_{\rm eff}=d-\gamma_V$, which approaches unity. These exponents describe volume growth with areal radius and do not establish a reduction of spatial dimension or a field-theory hyperscaling-violation exponent. At large proper radial distance, the volume grows exponentially
		according to \cref{proper_radius_volume_asymptotic}.
	}
	\label{fig:radial_volume_scaling}
\end{figure}

To describe the departure from Euclidean volume scaling and retain a comparison with hyperscaling, we introduce
\begin{align}
	\theta_{\rm eff}(R)
	\equiv d-\gamma_V(R)~,
	\label{theta_eff_definition}
\end{align}
which approaches zero at small areal radius and one at large areal radius. The crossover of both quantities is shown in \cref{fig:radial_volume_scaling} for $d=2,3,4,5$. Since $\theta_{\rm eff}(R)$ varies with radius, it characterizes a logarithmic slope rather than an exact power law throughout the crossover, while its asymptotic value $\theta_{\rm eff}\to1$ describes the leading $V_k(R)\propto R^{d-1}$ behaviour.

The interpretation of $\theta_{\rm eff}$ is therefore an analogy with hyperscaling violation at the level of the radial volume measure. In a theory with $n$ spatial dimensions, dynamical exponent $z$ and hyperscaling-violation exponent $\theta$, the thermal entropy density scales as $s_{\rm th}(T)\propto T^{(n-\theta)/z}$ in the corresponding scaling regime \cite{Dong:2012se}. Establishing this interpretation for the present construction would require a specified field theory or holographic model together with the scaling of its physical observables, and the configuration-space dimension $d$ has not been identified with the spatial dimension $n$ of a putative dual theory.

The results established here are the asymptotic relation between volume and boundary area in \cref{asymptotic_volume_area_ratio} and its contribution to the semiclassical count in \cref{radial_volume_semiclassical_count}, without an accompanying determination of an entropy area law, a spectral dimension or a holographic hyperscaling-violation exponent.

\subsection{Isotropic harmonic oscillator}
\label{sec:harmonic_oscillator}

We now study the spectrum of a confined particle in order to determine how the radial deformation affects the additional degeneracy of the ordinary oscillator, beyond that required by rotational symmetry. We consider $d\geq2$ and $k\geq0$, with $k=0$ understood as the undeformed limit, and introduce the central potential
\begin{align}
	V(r)=\frac12m\omega^2r^2~,
	\qquad \omega>0~.
	\label{osc_potential}
\end{align}
Since $r^2=\sum_i(x^i)^2$, this choice describes quadratic confinement in the position observables introduced above, whereas a potential quadratic in the proper radial distance $\tau_k$ would define a different physical system.

To specify the dynamics, we take the selected self-adjoint frame momenta to be the physical momentum observables and impose the usual quadratic relation between their second moments and the kinetic energy. For a normalized state in $\mathcal D(\mathfrak q_P)$, the prescription in \cref{initial_frame_hamiltonian_form} then reads
\begin{align}
	\mathfrak q_P[\psi]
	=\frac{1}{2m}\sum_{i=1}^d\|\widehat P_i\psi\|_k^2
	=\frac{1}{2m}\sum_{i=1}^d
	\left[(\Delta P_i)^2+\langle\widehat P_i\rangle_k^2\right]~.
	\label{osc_kinetic_momentum_postulate}
\end{align}
This is an additional dynamical assumption that connects the momentum observables entering the uncertainty relations to the energy of the particle, with no further scalar contribution added to their squared sum. It therefore gives a physical motivation for the frame prescription without requiring the deformation algebra alone to determine the Hamiltonian.

The oscillator is defined by the energy form
\begin{align}
	\mathfrak q_{\rm osc}[\psi]
	&=\mathfrak q_P[\psi]
	+\frac{m\omega^2}{2}
	\int_{\mathbb R^d}d\mu_k\,r^2|\psi|^2~,
	\label{osc_form}\\
	\mathcal D(\mathfrak q_{\rm osc})
	&=\mathcal D(\mathfrak q_P)
	\cap\left\{\psi\in\mathcal H_k
	\,\middle|\,r\psi\in\mathcal H_k\right\}~.
	\label{osc_form_domain}
\end{align}
This densely defined, closed and non-negative form determines a unique non-negative self-adjoint Hamiltonian $\widehat H_{\rm osc}^{(P)}$~\cite{Kato:1995}, which generates unitary time evolution and preserves rotational symmetry. Since the potential is regular at the center, the radial boundary conditions remain those selected in \cref{subsec_global_frame_hamiltonian}, with no independent interaction introduced there. The uniqueness of this Hamiltonian follows after specifying \cref{osc_form}, while the geometric prescription introduced in \cref{initial_geometric_hamiltonian_form} also gives a non-negative self-adjoint Hamiltonian, respects rotations and has the same classical kinetic energy. The distinction between these prescriptions can be expressed directly through the curvature of the selected radial geometry. Combining \cref{direct_radial_scalar_curvature,frame_vs_laplace_hamiltonian_local} gives the exact local identity
\begin{align}
	\widehat H_P^{\rm loc}
	=-\frac{\hbar^2}{2m}\Delta_g
	+\frac{\hbar^2}{8m}
	\left[\mathcal R+(d-1)k^2\right]~,
	\qquad \text{on }\mathcal D_0~.
	\label{osc_frame_curvature_coupling}
\end{align}
Thus the chosen frame prescription is equivalent, at the level of its local action, to a definite curvature coupling and an additive constant relative to the scalar Laplacian. Such curvature contributions also arise when comparing quantum kinetic prescriptions on curved configuration spaces~\cite{Ballesteros:2011}, although their coefficients are not fixed by rotational symmetry or the classical limit. In the present model, the position dependence of $\mathcal R$ implies that this difference can affect energy separations, providing a direct spectral test of the assumption in \cref{osc_kinetic_momentum_postulate}.

To obtain the radial spectrum, introduce the dimensionless variables
\begin{align}
	a_{\rm ho}&=\sqrt{\frac{\hbar}{m\omega}}~,
	&\kappa&=ka_{\rm ho}~,
	&\xi&=\frac{\tau_k}{a_{\rm ho}}~,
	\label{osc_small_parameter}\\
	\rho_\kappa(\xi)
	&\equiv\frac{r(a_{\rm ho}\xi)}{a_{\rm ho}}
	=\frac{e^{\kappa\xi}-1}{\kappa}~,
	\label{osc_dimensionless_radius}
\end{align}
where $\rho_0(\xi)=\xi$ by continuity. The radial transformation in \cref{radial_unitary_flattening_map}, followed by the normalization-preserving rescaling $v(\xi)=a_{\rm ho}^{1/2}u(a_{\rm ho}\xi)$, places each angular sector in $L^2((0,\infty),d\xi)$. Using \cref{partial_wave_frame_hamiltonian}, the dimensionless frame Hamiltonian becomes
\begin{align}
	\mathsf h_{\kappa,\ell}
	=\frac12\left[
	-\frac{d^2}{d\xi^2}
	+\frac{\nu_\ell^2-\frac14}{\rho_\kappa(\xi)^2}
	-\frac{\kappa(d-1)}{2\rho_\kappa(\xi)}
	+\rho_\kappa(\xi)^2
	\right]~,
	\label{osc_radial_exact}
\end{align}
where $\nu_\ell=\ell+(d-2)/2$ and
$E^{(P)}=\hbar\omega\,\varepsilon^{(P)}$. The center condition is inherited from \cref{osc_form}, while the wavefunction must be normalizable at infinity. For $\kappa>0$, the confining contribution grows as
\begin{align}
	\frac12\rho_\kappa(\xi)^2
	\sim\frac{e^{2\kappa\xi}}{2\kappa^2}
	\qquad (\xi\to\infty)~,
	\label{osc_large_tau}
\end{align}
while the inverse-radius terms vanish, giving a discrete bound-state spectrum that can be determined from \cref{osc_radial_exact} at finite deformation.

For small $\kappa$, the coefficient functions have the fixed-$\xi$ expansions
\begin{align}
	\rho_\kappa(\xi)
	&=\xi+\frac{\kappa}{2}\xi^2
	+O(\kappa^2\xi^3)~,
	\nonumber\\
	\rho_\kappa(\xi)^{-1}
	&=\xi^{-1}-\frac{\kappa}{2}
	+O(\kappa^2\xi)~,
	\nonumber\\
	\rho_\kappa(\xi)^{-2}
	&=\xi^{-2}-\frac{\kappa}{\xi}+O(\kappa^2)~.
	\label{osc_inverse_expansion}
\end{align}
The undeformed radial Hamiltonian and the coefficient of its first-order correction are therefore
\begin{align}
	\mathsf h_{0,\ell}
	&=\frac12\left[
	-\frac{d^2}{d\xi^2}
	+\frac{\nu_\ell^2-\frac14}{\xi^2}
	+\xi^2\right]~,
	\label{osc_H0}\\
	\mathsf h_{1,\ell}
	&=\frac12\left[
	\xi^3
	-\frac{\ell(\ell+d-2)+\frac14(d-1)^2}{\xi}
	\right]~.
	\label{osc_H1}
\end{align}
The cubic term arises from expressing the areal-radius potential \cref{osc_potential} in proper radial distance, while the inverse-radius term combines the expansion of the angular and radial-measure contributions with the scalar term in the frame kinetic energy.

Since \cref{osc_inverse_expansion} is not uniform at large $\xi$, we use it to expand individual energy levels at fixed radial and angular quantum numbers. For fixed $n$ and $\ell$, the squared undeformed radial wavefunctions decay as a power of $\xi$ multiplied by $e^{-\xi^2}$, so the region $\kappa\xi\gtrsim1$ is exponentially suppressed in the expectation values entering the first-order energy shifts. The undeformed energies are
\begin{align}
	E_{n\ell}^{(0)}
	=\hbar\omega\left(2n+\ell+\frac d2\right)~,
	\qquad n=0,1,2,\ldots~,
	\label{osc_E0}
\end{align}
and first-order perturbation theory gives
\begin{align}
	E_{n\ell}^{(P)}
	&=\hbar\omega\left[
	2n+\ell+\frac d2
	+\kappa C_{n\ell}^{(d)}+O(\kappa^2)
	\right]~,
	\label{osc_general_shift}\\
	C_{n\ell}^{(d)}
	&=\frac12\left[
	\langle\xi^3\rangle_{n\ell}^{(0)}
	-\left(\ell(\ell+d-2)+\frac{(d-1)^2}{4}\right)
	\langle\xi^{-1}\rangle_{n\ell}^{(0)}
	\right]~.
	\label{osc_shift_coefficient}
\end{align}
Although \cref{osc_E0} is degenerate, the central perturbation does not mix different angular sectors and acts identically on their angular degeneracy labels. Within a shell of fixed $N=2n+\ell$, each allowed $\ell$ determines a unique $n$, so \cref{osc_shift_coefficient} directly gives the first-order shifts without additional diagonalization.

The expectation values in normalized eigenstates of \cref{osc_H0} are
\begin{align}
	\langle\xi^s\rangle_{n\ell}^{(0)}
	=\frac{n!}{\Gamma(n+\nu_\ell+1)}
	\int_0^\infty dz\,
	z^{\nu_\ell+s/2}e^{-z}
	\left[L_n^{\nu_\ell}(z)\right]^2~,
	\label{osc_general_moment}
\end{align}
where $z=\xi^2$ and $L_n^{\nu_\ell}$ denotes an associated Laguerre polynomial. The convergence condition $s>-2\nu_\ell-2$ is satisfied by both moments in \cref{osc_shift_coefficient} for $d\geq2$, and for the entire $n=0$ family the integrals give
\begin{align}
	C_{0\ell}^{(d)}
	=\left(\ell+\frac{d-1}{4}\right)
	\frac{\Gamma\left(\ell+\frac{d-1}{2}\right)}
	{\Gamma\left(\ell+\frac d2\right)}~.
	\label{osc_n0_coefficient}
\end{align}
This coefficient is positive for $d\geq2$, so positive deformation raises the energies of these states at first order. In particular, the ground-state energy becomes
\begin{align}
	E_{00}^{(P)}
	=\hbar\omega\left[
	\frac d2
	+\kappa\frac{d-1}{4}
	\frac{\Gamma\left(\frac{d-1}{2}\right)}{\Gamma(d/2)}
	+O(\kappa^2)\right]~,
	\label{osc_ground_shift_general}
\end{align}
which in three dimensions reduces to
\begin{align}
	E_{00}^{(P)}
	=\hbar\omega\left[
	\frac32+\frac{\kappa}{\sqrt{\pi}}+O(\kappa^2)
	\right]~.
	\label{osc_ground_shift_3d}
\end{align}

The excited spectrum distinguishes states that share the same undeformed energy but have different radial and orbital quantum numbers. Writing $C_{n\ell}=C_{n\ell}^{(3)}$ for the frame prescription, the first few coefficients are
\begin{align}
	C_{00}&=\frac{1}{\sqrt{\pi}}~,
	&C_{01}&=\frac{2}{\sqrt{\pi}}~,
	\nonumber\\
	C_{10}&=\frac{37}{6\sqrt{\pi}}~,
	&C_{02}&=\frac{8}{3\sqrt{\pi}}~,
	\nonumber\\
	C_{11}&=\frac{41}{5\sqrt{\pi}}~,
	&C_{03}&=\frac{16}{5\sqrt{\pi}}~.
	\label{osc_low_coefficients}
\end{align}
Rotational symmetry preserves the $(2\ell+1)$ magnetic degeneracy exactly, while the separate dependence on $n$ and $\ell$ lifts the additional degeneracy between the angular sectors displayed in \cref{osc_low_coefficients}. For the $N=2$ shell, the states $(n,\ell)=(1,0)$ and $(0,2)$ acquire the splitting
\begin{align}
	E_{10}^{(P)}-E_{02}^{(P)}
	=\frac{7}{2\sqrt{\pi}}\,\kappa\hbar\omega
	+O(\kappa^2\hbar\omega)~,
	\label{osc_N2_splitting}
\end{align}
so the radially excited state lies above the state with higher orbital angular momentum for sufficiently small positive $\kappa$. The unequal coefficients $C_{11}$ and $C_{03}$ likewise split the $N=3$ shell, showing that the additional shell degeneracy of the ordinary oscillator is not preserved by this deformation of its dynamics.

\begin{figure}[t]
	\centering
	\includegraphics[width=\textwidth]{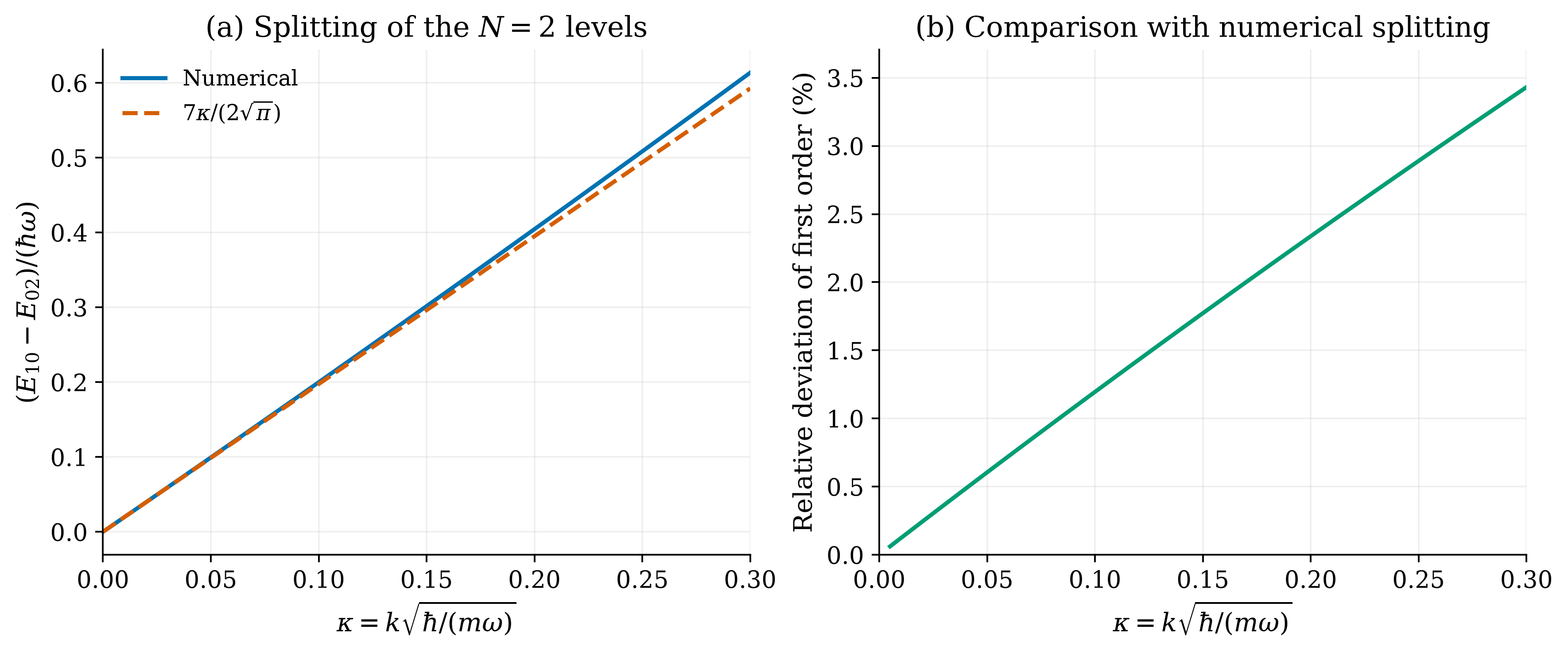}
	\caption{
		Splitting of the $N=2$ oscillator shell for the three-dimensional frame Hamiltonian, with $\kappa=k\sqrt{\hbar/(m\omega)}$. The left panel compares the numerical splitting obtained from \cref{osc_radial_exact} with the first-order result $\frac{E_{10}^{(P)}-E_{02}^{(P)}}{\hbar\omega} \simeq\frac{7\kappa}{2\sqrt{\pi}}$ in \cref{osc_N2_splitting}, while the right panel shows the relative deviation $100|\Delta\varepsilon_{\rm num} -\Delta\varepsilon_{\rm first}|/ |\Delta\varepsilon_{\rm num}|$ for $\kappa>0$, where $\Delta\varepsilon=(E_{10}^{(P)}-E_{02}^{(P)})/(\hbar\omega)$. Both splittings vanish at $\kappa=0$, where the relative deviation is undefined. The calculation uses the areal-radius potential \cref{osc_potential} and the Friedrichs condition at the center.
	}
	\label{fig:oscillator_splitting}
\end{figure}

To compare \cref{osc_N2_splitting} with the spectrum at finite deformation, we solve \cref{osc_radial_exact} numerically in three dimensions, imposing the Friedrichs condition at the center and a vanishing reduced wavefunction at a distant outer endpoint. Centred finite differences on $0\leq\xi\leq12$, with $4000$ and $8000$ intervals, allow extrapolation to remove the leading quadratic discretization error. Checks at $\kappa=0,0.1,0.3$ using finer grids and an outer endpoint at $\xi=15$ change the relevant dimensionless energies by less than $10^{-8}$, well below the scale resolved in \cref{fig:oscillator_splitting}. The numerical splitting agrees with the first-order result at small $\kappa$, while the relative deviation reaches approximately $3.43\%$ at $\kappa=0.3$. The expansion concerns fixed levels at small $\kappa$, since its applicability also depends on the spatial extent of the state. Using $\langle\xi^2\rangle_{n\ell}^{(0)}=2n+\ell+d/2$, the coordinate expansion requires $\kappa\sqrt{2n+\ell+d/2}\ll1$ over the characteristic extent of an excited state. Consequently, \cref{osc_general_shift} and the numerical comparison in \cref{fig:oscillator_splitting} do not establish a uniform approximation for arbitrarily high oscillator levels.

We finally compare these predictions with the geometric kinetic prescription, keeping the same potential, radial geometry and Friedrichs condition at the center. Let $E_{n\ell}^{(g)}$ denote the oscillator energies obtained by adding \cref{osc_potential} to the Hamiltonian $\widehat H_g^F$ defined in \cref{initial_geometric_hamiltonian_form}. The quantum potential in \cref{frame_vs_laplace_hamiltonian_local} gives, at fixed $n$ and $\ell$,
\begin{align}
	E_{n\ell}^{(g)}-E_{n\ell}^{(P)}
	=\kappa\hbar\omega\,
	\frac{(d-1)^2}{4}
	\langle\xi^{-1}\rangle_{n\ell}^{(0)}
	+O(\kappa^2\hbar\omega)~.
	\label{osc_ordering_energy_difference}
\end{align}
The expectation value is state dependent, so this difference cannot in general be absorbed into a change of the energy zero. For example, the three-dimensional ground-state energy becomes
\begin{align}
	E_{00}^{(g)}
	=\hbar\omega\left[
	\frac32+\frac{3\kappa}{\sqrt{\pi}}+O(\kappa^2)
	\right]~,
	\label{osc_geometric_ground_shift_3d}
\end{align}
while a comparison of the split shell tests the effect on energy separations directly. In three dimensions, \cref{osc_general_moment} gives
\begin{align}
	\langle\xi^{-1}\rangle_{10}^{(0)}
	&=\frac{5}{3\sqrt{\pi}}~,
	&\langle\xi^{-1}\rangle_{02}^{(0)}
	&=\frac{16}{15\sqrt{\pi}}~,
	\label{osc_N2_inverse_moments}
\end{align}
and hence
\begin{align}
	E_{10}^{(g)}-E_{02}^{(g)}
	=\frac{41}{10\sqrt{\pi}}\,\kappa\hbar\omega
	+O(\kappa^2\hbar\omega)~.
	\label{osc_N2_geometric_splitting}
\end{align}
The expressions described in \cref{osc_N2_splitting,osc_N2_geometric_splitting} show that both prescriptions lift the $N=2$ shell degeneracy at first order and place the radially excited state above the state with higher orbital angular momentum, while preserving the magnetic degeneracy. For the same values of $k$, $m$ and $\omega$, the geometric prescription gives a first-order $N=2$ splitting that is $41/35$ times the frame result, so the two prescriptions differ in an energy separation rather than only in the choice of energy zero.

The oscillator therefore gives a definite dynamical prediction once the momentum observables, kinetic-energy prescription, confining potential and center condition have been specified. The lifting and ordering of the $N=2$ shell survive the comparison between the two prescriptions considered here, whereas the magnitude of the splitting distinguishes them. This agreement does not establish a universal splitting for every quantization of the radial geometry, since rotational symmetry protects the magnetic degeneracy but does not fix the relative energies of different angular sectors.

\section{Discussion}
\label{dis}

We have developed a rotationally covariant quantum model that connects a nonadditive radial displacement law to momentum observables, localization bounds and a calculable bound-state spectrum. We retain the radial derivative and integration rule suggested by the scalar calculus, combine the radial measure with the usual spherical volume factor, and choose the momentum frame so that its associated geometric volume measure agrees with this spatial measure. Within the class of frames considered in this work, these conditions leave the tangential derivative coefficients undeformed and fix the spatial construction. The model connects the radial deformation to the momentum algebra and to the splitting of harmonic-oscillator energy levels with an explicitly chosen quantum kinetic-energy prescription.

The distinction between radial and tangential motion is central to this construction. Earlier nonadditive displacement models connect position-dependent commutators with position-dependent effective masses \cite{CostaFilho:2011}, while related constructions link a common scalar deformation of all momentum components to effective-mass systems and curved configuration spaces, including examples with a deformation factor linear in radius \cite{Quesne:2004}. The direct completion instead combines the prescribed radial response with undeformed tangential coefficients, which changes the full momentum algebra, geometric measure and kinetic operator. This comparison shows why agreement at the level of a scalar radial derivative is insufficient to identify two multidimensional quantum models. Their treatment of angular motion carries additional physical information, which becomes accessible through states with different orbital angular momenta.

The induced geometry makes this additional information explicit. Although the radial derivative can be written as ordinary differentiation with respect to proper radial distance, the angular part of the metric retains a nontrivial dependence on that distance. The full configuration space therefore remains curved, with position-dependent negative curvature away from the center and an approach to constant negative curvature at large distance. A consistent change of coordinates preserves these properties and the predictions of the specified Hamiltonian. The distinguished center is consequently part of the effective background, while the selected frame defines the local momentum components used to describe the particle. Rotational covariance is preserved throughout, without requiring ordinary spatial translation invariance.

The construction of self-adjoint Cartesian frame momenta establishes that these components are well-defined quantum observables, including their global action in the weighted Hilbert space. Their noncommutativity is determined by the spatial variation of the frame and takes the form of an angular momentum contribution with a radius-dependent coefficient. Consequently, a single Cartesian momentum representation cannot diagonalize all frame components simultaneously, although individual components retain their spectral resolutions and the Hamiltonian can be analyzed in angular momentum sectors. Canonical momenta with the ordinary Heisenberg algebra remain available in the same Hilbert space, making clear that the modified algebra belongs to the selected mechanical components and is not imposed by the integration measure alone.

For these frame observables, the uncertainty analysis yields a positive correction governed by the mean areal radius of the state relative to the distinguished center. Combining the directional inequalities gives a bound for the root-mean-square averaged uncertainties without assuming a rotationally symmetric wavefunction. The result therefore applies to angularly structured states as well as to radial states within the test domain used in the derivation. Its physical content is an increased lower bound on the combined coordinate-position and frame-momentum fluctuations as the state samples larger radial scales. The mean radius remains distinct from the spatial width, and arbitrarily narrow smooth wave packets remain admissible with increasing momentum uncertainties. The construction thus gives a position-dependent uncertainty bound without imposing a universal minimum coordinate length, while the optimality and saturation of the derived bound remain separate questions.

Connecting this kinematic result to dynamics requires a Hamiltonian whose definition includes the admissible behaviour of wavefunctions. We have provided such a definition by constructing a non-negative self-adjoint kinetic Hamiltonian from the squared norms of the frame momenta. The radial spectral problem is then determined by this Hamiltonian and its center condition. This distinction is necessary because the first-order radial momentum retains the familiar obstruction associated with a half-line, which is already present in undeformed radial quantum mechanics. The phase modes inherited from the scalar composition law therefore do not supply a spectral basis for a self-adjoint radial momentum. The Hamiltonian construction provides the appropriate basis for radial dynamics and makes the subsequent spectral calculations independent of that first-order obstruction.

The isotropic harmonic oscillator demonstrates the dynamical content of the model through energy differences between states that are degenerate in the ordinary theory. For a confining potential quadratic in the areal radius, we obtained the exact radial Hamiltonian and the leading corrections to fixed energy levels at small positive deformation. The ground-state energy increases at first order, while the excited-state corrections depend separately on radial and orbital quantum numbers. In the three-dimensional shell examined numerically, the first radially excited state with zero angular momentum lies above the state with $n=0$ and $\ell=2$, although their undeformed energies coincide. The magnetic degeneracy within each angular sector remains intact, so the splitting distinguishes the additional degeneracy of the ordinary oscillator from the degeneracy protected by rotations.

The comparison in \cref{fig:oscillator_splitting} confirms the leading splitting in the small-deformation regime and quantifies its departure from the numerical result at finite deformation for the selected pair of levels. This supplies a concrete test of the analytical expansion with a specified center condition and a controlled undeformed limit. The approximation is intended for fixed levels, since increasingly excited states extend into regions where the radial deformation is larger. The spectral result also depends on maintaining the stated confining potential when changing radial coordinates, because a potential quadratic in proper distance would describe a different system. These features identify the physical assumptions behind the splitting and show how a radial deformation can affect shell structure while preserving rotational symmetry.

The comparison with the geometric kinetic prescription further demonstrates that the spatial metric alone does not fix the quantum spectrum. Although the two prescriptions have the same classical kinetic energy, their local quantum expressions differ by an explicit scalar potential, producing different energy shifts for the same confining potential. For the same deformation and oscillator parameters, the leading ground-state shift obtained from the geometric prescription in three dimensions is three times that obtained from the frame prescription. The difference also depends on the state, so it cannot generally be absorbed into a common choice of energy zero. This turns the ordering freedom into a quantitative comparison and clarifies why the uncertainty bound alone cannot determine the oscillator spectrum. The predicted shifts and splittings belong to the specified combination of spatial frame, kinetic prescription and confining potential.

The singular center provides another place where the physical specification of the model matters. Our kinetic-energy construction selects a definite radial boundary prescription, with no additional center parameter left undetermined in the Hamiltonian used for the oscillator. For the same local radial expression, alternative self-adjoint conditions are available only in the zero angular momentum sectors in two and three spatial dimensions, whereas the remaining sectors admit no independent center condition. This result locates precisely where a different microscopic core could enter through boundary data. It also separates the existence of a geometric singularity from an ambiguity in the dynamics, since the stated Hamiltonian already supplies a definite evolution and spectrum.

The geometric measure has an additional consequence for the number of states available in a bounded region. The volume enclosed by a centred sphere has Euclidean leading behaviour at small areal radius, but becomes proportional to the boundary area at large areal radius, with the deformation length setting the proportionality scale. For a particle confined to the ball, this volume supplies the spatial factor in the leading high-energy semiclassical count at fixed ball radius, while the energy exponent retains its ordinary dependence on the number of spatial dimensions. The result therefore connects the measure used to normalize wavefunctions with a concrete counting problem, without identifying the volume integral with the complete spectrum of the noncompact system.

The comparison with proper radial distance is essential to the interpretation of this volume law. At large proper radial distance, the enclosed volume grows exponentially, consistently with the asymptotic negative curvature, even though its leading power in the areal radius is reduced by one. The effective exponent introduced in the volume analysis describes this geometric crossover and provides a limited analogy with hyperscaling violation. In field theory or holography, such an exponent must instead be established from the scaling of specified physical observables \cite{Dong:2012se}. The present calculation does not establish a reduction of spatial dimension, an entropy area law or a holographic dual, but it identifies the geometric scaling that any proposed interpretation would need to relate to thermodynamics or other observables.

A physical realization of the construction would need to explain how the preferred frame and the distinguished center arise, and which kinetic prescription governs the probe. An effective radial medium or a localized defect could provide a setting in which these questions can be addressed by deriving the particle Hamiltonian from more microscopic degrees of freedom. Such a derivation would also identify the meaning of the deformation length, since the present spatial kinematics determines only one dimensionful combination of the nonextensivity parameter and the reference scale. Consequently, measurements of the calculated spectral effects could constrain that combination within a specified realization, while an independent interpretation of the statistical parameter would require further physical input.

Several extensions follow directly from the results obtained here. Resolving the singular center with a smooth core would allow the sensitivity of low-energy levels and radial scattering to the short-distance structure to be examined, including whether a given microscopic construction approaches the center prescription adopted in this work. Transition matrix elements and scattering observables would provide further dynamical tests of the same momentum frame, while a systematic comparison with other tangential responses could identify which spectral features distinguish different completions of the scalar law. These directions build on the explicit Hamiltonian and oscillator calculation, whose main contribution is to turn the proposed radial kinematics into definite quantum predictions with identifiable geometric and dynamical assumptions.

\section*{Acknowledgment}
We thank Rajesh Karmakar for helpful discussions. We are also grateful to ChatGPT (OpenAI) for assistance with manuscript preparation, including language editing and the exploration of mathematical derivations relevant to this work. This work has been supported by NSFC, China (Grant No. 12275166 and No. 12675074).

\appendix

\section{Weak frame derivatives and the form core}
\label{app_weak_domain_form_core}

This appendix establishes \cref{D0_form_core_statement} for $d\geq2$ by showing that every state in $\mathcal D(\mathfrak q_P)$ can be approximated by smooth wavefunctions with compact support away from the center, with convergence in both norm and kinetic energy. The argument uses the unitary flows and self-adjoint generators defined in \cref{weighted_flow_unitary,Stone_generator_definition,Stone_generator_domain}, together with the weak derivative relation that follows from their action on smooth test functions. For $\psi\in\mathcal D(\widehat P_i)$, write $Y_i\psi=(i/\hbar)\widehat P_i\psi$, so the norm introduced in \cref{frame_form_norm_equivalent} becomes
\begin{align}
	\|\psi\|_{P,1}^2
	&=
	\|\psi\|_k^2
	+
	\frac{\hbar^2}{2mE_*}
	\sum_{i=1}^d\|Y_i\psi\|_k^2~,
	\qquad
	\psi\in\mathcal D(\mathfrak q_P)~,
	\label{appendix_equivalent_form_norm}
\end{align}
where the arbitrary reference energy $E_*>0$ makes the two contributions dimensionally consistent without changing the convergence required below.

The positive weight $w_k$ and the frame coefficients $E_i{}^a$ are globally Lipschitz, so their product is locally Lipschitz and its divergence is well defined in the weak sense on the entire space. Consequently, the result obtained away from the center in \cref{direct_radial_connection_term} extends to the full space in the weak sense, with the weighted divergence coefficient given almost everywhere by
\begin{align}
	a_i(x)
	&\equiv
	\frac{1}{w_k(x)}
	\partial_a\!\left(w_k(x)E_i{}^a(x)\right)
	=
	k(d-1)n_i(x)~.
	\label{appendix_weighted_divergence_ai}
\end{align}
The value assigned at the center does not affect any of the integrals, while the local Lipschitz regularity ensures that the divergence contains no additional contribution concentrated there.

To extend the infinitesimal action from $\mathcal D_0$ to smooth test functions that may be nonzero at the center, consider the density factor in \cref{flow_density_factor}. The change of weighted volume along the inverse flow gives
\begin{align}
	\rho_i(t,x)
	&=
	\exp\!\left[
	-\int_0^t a_i\!\left(\Phi_i^{-s}(x)\right)ds
	\right]~,
	\label{appendix_weighted_liouville_formula}
\end{align}
for almost every $x$, and the bound $M_i=\|a_i\|_\infty=k(d-1)$ implies
\begin{align}
	\left|
	\frac{\rho_i(t,x)^{1/2}-1}{t}
	\right|
	&\leq
	\frac{M_i}{2}e^{M_i|t|/2}~.
	\label{appendix_density_derivative_bound}
\end{align}
Since $a_i$ is continuous away from the center, this difference quotient converges almost everywhere to $-a_i/2$ as $t\to0$. For $\varphi\in C_c^\infty(\mathbb R^d)$, the composition difference quotient converges to $-X_i\varphi$, and all supports remain in a common bounded region for sufficiently small $|t|$. The bound in \cref{appendix_density_derivative_bound} therefore allows dominated convergence to be applied to \cref{weighted_flow_unitary}, giving
\begin{align}
	\left.
	\frac{d}{dt}U_i(t)\varphi
	\right|_{t=0}
	&=
	-\left(X_i+\frac{a_i}{2}\right)\varphi
	\qquad \text{in }\mathcal H_k~,
	\label{appendix_test_function_derivative}\\
	\widehat P_i\varphi
	&=
	-i\hbar\left(X_i+\frac{a_i}{2}\right)\varphi~,
	\qquad
	\varphi\in C_c^\infty(\mathbb R^d)~.
	\label{appendix_local_action_full_tests}
\end{align}
Thus these full-space test functions belong to $\mathcal D(\widehat P_i)$ even when their support includes the center.

For $\psi\in\mathcal D(\widehat P_i)$, self-adjointness  and \cref{appendix_local_action_full_tests} now imply
\begin{align}
	\langle\varphi,Y_i\psi\rangle_k
	&=
	-\left\langle
	\left(X_i+\frac{a_i}{2}\right)\varphi,\psi
	\right\rangle_k~,
	\qquad
	\varphi\in C_c^\infty(\mathbb R^d)~.
	\label{appendix_full_weak_identity}
\end{align}
Because the test functions need not vanish at the center, this identity determines the weak frame derivative across the entire space. It also extends to compactly supported Lipschitz test functions by smoothing them, since the smoothed functions converge uniformly, their gradients remain uniformly bounded and a sequence of those gradients converges almost everywhere.

For $\psi\in\mathcal D(\mathfrak q_P)$, the identity holds for every frame component and can be used to recover the Cartesian weak derivatives. Taking $\varphi=\zeta/w_k$ in \cref{appendix_full_weak_identity} and using $a_i=\partial_aE_i{}^a+X_i(\log w_k)$ gives
\begin{align}
	\int_{\mathbb R^d}d^dx\,
	\zeta^*\left(Y_i\psi-\frac{a_i}{2}\psi\right)
	&=
	-\int_{\mathbb R^d}d^dx\,
	\psi\,\partial_a\!\left(E_i{}^a\zeta^*\right)~,
	\label{appendix_ordinary_distribution_frame_derivative}
\end{align}
for every compactly supported Lipschitz function $\zeta$. For a fixed Cartesian index $a$, choose the test function in the $i$th identity to be $\zeta_i=(E^{-1})_a{}^i\zeta$, with $\zeta\in C_c^\infty(\mathbb R^d)$. These test functions are again Lipschitz with compact support, so summing over $i$ and using $E_i{}^b(E^{-1})_a{}^i=\delta_a{}^b$ yields the following identity in the weak sense on $\mathbb R^d$,
\begin{align}
	\partial_a\psi
	&=
	(E^{-1})_a{}^i
	\left(Y_i\psi-\frac{a_i}{2}\psi\right)~.
	\label{appendix_distributional_partial_from_frame}
\end{align}

On every bounded region, the weight is bounded above and below by positive constants, while $E$, $E^{-1}$ and $a_i$ are bounded. It follows from \cref{appendix_distributional_partial_from_frame} that $\psi$ and its Cartesian weak derivatives are locally square integrable, so $\psi\in H^1_{\mathrm{loc}}(\mathbb R^d)$. Conversely, the expression $Y_i\psi=E_i{}^a\partial_a\psi+a_i\psi/2$ bounds the frame derivatives in terms of the Cartesian derivatives. Consequently, for states supported in a common bounded region, convergence in the ordinary $H^1$ norm gives convergence in the frame form norm. To remove the tail at infinity, choose a smooth function $\chi$ on $[0,\infty)$ with $0\leq\chi\leq1$, $\chi(s)=1$ for $s\leq1$ and $\chi(s)=0$ for $s\geq2$, and set
\begin{align}
	\chi_R(x)
	&=
	\chi\!\left(\frac{\tau_k(r)}{R}\right)~,
	\qquad R>0~.
	\label{appendix_infinity_cutoff_definition}
\end{align}
Here $R$ has dimensions of length, and the additive coordinate in \cref{radial_additive_coordinate} satisfies
\begin{align}
	X_i\tau_k
	&=
	n_i~,
	&
	X_i\chi_R
	&=
	\frac{1}{R}
	\chi'\!\left(\frac{\tau_k(r)}{R}\right)n_i~.
	\label{appendix_infinity_cutoff_derivative}
\end{align}
In particular, $\chi_R$ has compact support and $\|X_i\chi_R\|_\infty\leq\|\chi'\|_\infty/R$.

The flow action gives $U_i(t)(\chi_R\psi) =(\chi_R\circ\Phi_i^{-t})U_i(t)\psi$, while the difference quotient of $\chi_R\circ\Phi_i^{-t}$ converges to $-X_i\chi_R$ and is bounded by $\|X_i\chi_R\|_\infty$. Dominated convergence and the strong continuity of $U_i(t)$ therefore show that $\chi_R\psi\in\mathcal D(\widehat P_i)$ whenever $\psi\in\mathcal D(\widehat P_i)$, with
\begin{align}
	Y_i(\chi_R\psi)
	&=
	\chi_RY_i\psi+(X_i\chi_R)\psi~.
	\label{appendix_infinity_cutoff}
\end{align}
For $\psi\in\mathcal D(\mathfrak q_P)$, the first term converges to $Y_i\psi$ by dominated convergence, while the second has norm bounded by $\|\chi'\|_\infty\|\psi\|_k/R$. Together with $\chi_R\psi\to\psi$ in $\mathcal H_k$, this proves
\begin{align}
	\|\chi_R\psi-\psi\|_{P,1}
	&\longrightarrow0
	\qquad (R\to\infty)~.
	\label{appendix_tail_approximation}
\end{align}

For fixed $R$, the state $\psi_R=\chi_R\psi$ has compact support and belongs to $H^1(\mathbb R^d)$, so smoothing by convolution gives $\varphi_n\in C_c^\infty(\mathbb R^d)$, supported in a common bounded ball, with $\varphi_n\to\psi_R$ in $H^1$ \cite{AdamsFournier:2003}. These smooth functions belong to every $\mathcal D(\widehat P_i)$ by \cref{appendix_local_action_full_tests}, and the boundedness of the coefficients on their common support therefore gives
\begin{align}
	\|\varphi_n-\psi_R\|_{P,1}
	&\longrightarrow0
	\qquad (n\to\infty)~.
	\label{appendix_smooth_approximation}
\end{align}

The center can now be removed from each fixed smooth function with arbitrarily small gradient cost, which is the zero $H^1$ capacity of a point for $d\geq2$ \cite{Mazya:2011}. Choose smooth radial cutoffs $0\leq\eta_\epsilon\leq1$ which vanish in a neighborhood of the origin, equal one outside a fixed ball, approach one at every nonzero point as $\epsilon\to0$ and satisfy
\begin{align}
	\|\nabla\eta_\epsilon\|_{L^2(\mathbb R^d)}
	&\longrightarrow0
	\qquad (\epsilon\to0)~.
	\label{appendix_center_cutoff_bound}
\end{align}
An ordinary shrinking radial cutoff has this property for $d>2$, while a logarithmic profile is required in $d=2$. For each fixed $\varphi_n$, the identity
\begin{align}
	\nabla(\eta_\epsilon\varphi_n-\varphi_n)
	&=
	(\eta_\epsilon-1)\nabla\varphi_n
	+
	\varphi_n\nabla\eta_\epsilon
	\label{appendix_center_cutoff_gradient}
\end{align}
shows that the gradient difference tends to zero in $L^2$, because the first term is controlled by dominated convergence and the second by the boundedness of $\varphi_n$ and \cref{appendix_center_cutoff_bound}. The wavefunction difference also tends to zero in $L^2$, so
\begin{align}
	\|\eta_\epsilon\varphi_n-\varphi_n\|_{H^1(\mathbb R^d)}
	&\longrightarrow0
	\qquad (\epsilon\to0)~.
	\label{appendix_center_approximation}
\end{align}
Since these products have support in a common bounded ball, the same convergence holds in the frame form norm, while $\eta_\epsilon\varphi_n\in\mathcal D_0$.

For any prescribed accuracy, the tail, smoothing and center-cutoff errors can therefore be made successively small, giving an approximation to every $\psi\in\mathcal D(\mathfrak q_P)$ by states in $\mathcal D_0$. Since the form in \cref{initial_frame_hamiltonian_form} is closed, these approximations establish
\begin{align}
	\overline{\mathcal D_0}^{\,\|\cdot\|_{P,1}}
	&=
	\mathcal D(\mathfrak q_P)~.
	\label{appendix_D0_form_core}
\end{align}
Thus the local form in \cref{local_frame_hamiltonian_form} closes to the global frame form, which identifies its Hamiltonian with the Friedrichs realization and justifies the center prescription used in \cref{subsec_global_frame_hamiltonian}.

\bibliographystyle{JHEP}

\bibliography{tsallis}

\end{document}

%% file: no-revtex.tex
\newcommand{\preprint}[1]{\begin{table}[t]  
             \begin{flushright}               
             {#1}                             
             \end{flushright}                 
             \end{table}}                     
\renewcommand{\title}[1]{\vbox{\center\LARGE{#1}}\vspace{5mm}}
\renewcommand{\author}[1]{\vbox{\center#1}\vspace{5mm}}
\newcommand{\address}[1]{\vbox{\center\em#1}}
\newcommand{\email}[1]{\vbox{\center\tt#1}\vspace{5mm}}

\renewcommand{\date}[1]{\vbox{\center#1}}



%% file: jm-tex-macros-public.tex
\usepackage{amssymb}
\usepackage{amsmath,bm}
\usepackage{amssymb}
\usepackage{graphicx}
\usepackage{amsfonts}         
\usepackage{fancybox}   

\usepackage{enumitem}

\usepackage{slashed}

\usepackage{epstopdf}
\usepackage[usenames,dvipsnames]{xcolor}
\usepackage[pagebackref, bookmarks={false}, pdfauthor={JD}, pdftitle={AbaniBariAacho}]{hyperref}
\hypersetup{colorlinks=true, linkcolor=darkblue, citecolor=red, filecolor=OliveGreen, urlcolor=blue, filebordercolor={.8 .8 1}, urlbordercolor={.8 .8 0}}
\usepackage{soul}
\setstcolor{Red}

\usepackage{wrapfig}

\definecolor{jazzberryjam}{rgb}{0.65, 0.04, 0.37}
\definecolor{lust}{rgb}{0.9, 0.13, 0.13}
\definecolor{sandybrown}{rgb}{0.96, 0.64, 0.38}
\definecolor{mountainmeadow}{rgb}{0.19, 0.73, 0.56}
\definecolor{glaucous}{rgb}{0.38, 0.51, 0.71}
\definecolor{chromeyellow}{rgb}{1.0, 0.65, 0.0}
\definecolor{emerald}{rgb}{0.31, 0.78, 0.47}
\definecolor{deepsaffron}{rgb}{1.0, 0.6, 0.2}
\definecolor{darkgreen}{rgb}{0,0.4,0}
\definecolor{darkred}{rgb}{0.4,0,0}
\definecolor{darkblue}{rgb}{0,0,0.4}
\definecolor{lightblue}{rgb}{.6,.6,0.9}

\definecolor{uglybrown}{rgb}{0.8,  0.7,  0.5}

\definecolor{palatinatepurple}{rgb}{0.41, 0.16, 0.38}
\definecolor{celebrationcolor}{rgb}{0.75,  0.0,  0.9}

 \usepackage{mdframed}

\usepackage{framed}
\definecolor{shadecolor}{rgb}{0.90,0.90,0.90}

\usepackage{tkz-euclide}

\usepackage{makeidx}
\usepackage{cite}
\usepackage{bm}
\usepackage{geometry}
\usepackage{braket}
\usepackage[margin=20pt,small]{caption}

\usepackage[toc]{appendix}

\usepackage{tikz}
\usetikzlibrary{calc,decorations.markings,arrows.meta,shapes.misc,decorations.pathmorphing,calc,bending}

\usetikzlibrary{arrows.meta,shapes.misc,decorations.pathmorphing,calc,bending}

\tikzset{
  branch point/.style={cross out,draw=black,fill=none,minimum size=2*(#1-\pgflinewidth),inner sep=0pt,outer sep=0pt}, 
  branch point/.default=5
}
\tikzset{
  branch cut/.style={
    decorate,decoration=snake,
    to path={
      (\tikztostart) -- (\tikztotarget) \tikztonodes
    },
    }
  }

\input epsf

\newlength{\extraspace}
\newlength{\extraspaces}
\def\be{\begin{equation}}
\def\ee{\end{equation}}

\newcommand{\bea}{\begin{eqnarray}}
\newcommand{\eea}{\end{eqnarray}}

\def\II{\relax{I\kern-.10em I}}

\def\IB{\relax{\rm I\kern-.18em B}}

\def\ID{\relax{\rm I\kern-.18em D}}
\def\IE{\relax{\rm I\kern-.18em E}}
\def\IF{\relax{\rm I\kern-.18em F}}
\def\IG{\relax\hbox{$\inbar\kern-.3em{\rm G}$}}
\def\IGa{\relax\hbox{${\rm I}\kern-.18em\Gamma$}}
\def\IH{\relax{\rm I\kern-.18em H}}
\def\II{\relax{\rm I\kern-.18em I}}
\def\IK{\relax{\rm I\kern-.18em K}}

\def\inbar{\,\vrule height1.5ex width.4pt depth0pt}

\def\IR{\mathbb{R}}

\def\lp10{\ell_p^{10}}
\def\lp11{\ell_p^{11}}
\def\R11{R_{11}}

\def\frac#1#2{{#1 \over #2}}

\newdimen\tableauside\tableauside=1.0ex
\newdimen\tableaurule\tableaurule=0.4pt
\newdimen\tableaustep
\def\phantomhrule#1{\hbox{\vbox to0pt{\hrule height\tableaurule width#1\vss}}}
\def\phantomvrule#1{\vbox{\hbox to0pt{\vrule width\tableaurule height#1\hss}}}
\def\sqr{\vbox{%
  \phantomhrule\tableaustep
  \hbox{\phantomvrule\tableaustep\kern\tableaustep\phantomvrule\tableaustep}%
  \hbox{\vbox{\phantomhrule\tableauside}\kern-\tableaurule}}}
\def\squares#1{\hbox{\count0=#1\noindent\loop\sqr
  \advance\count0 by-1 \ifnum\count0>0\repeat}}
\def\tableau#1{\vcenter{\offinterlineskip
  \tableaustep=\tableauside\advance\tableaustep by-\tableaurule
  \kern\normallineskip\hbox
    {\kern\normallineskip\vbox
      {\gettableau#1 0 }%
     \kern\normallineskip\kern\tableaurule}%
  \kern\normallineskip\kern\tableaurule}}
\def\gettableau#1 {\ifnum#1=0\let\next=\null\else
  \squares{#1}\let\next=\gettableau\fi\next}

 \def\eqnn#1{\xdef #1{(\secsym\the\meqno)}\writedef{#1\leftbracket#1}%
 \global\advance\meqno by1\wrlabeL#1}
 \def\eqna#1{\xdef #1##1{\hbox{$(\secsym\the\meqno##1)$}}
 \writedef{#1\numbersign1\leftbracket#1{\numbersign1}}%
 \global\advance\meqno by1\wrlabeL{#1$\{\}$}}
 \def\eqn#1#2{\xdef #1{(\secsym\the\meqno)}\writedef{#1\leftbracket#1}%
 \global\advance\meqno by1$$#2\eqno#1\eqlabeL#1$$}

\global\newcount\itemno \global\itemno=0

\def\itemaut#1{\global\advance\itemno by1\noindent\item{\the\itemno.}#1}

\def\({\left(}
\def\){\right)}

\def\lsim{\mathrel{\mathstrut\smash{\ooalign{\raise2.5pt\hbox{$<$}\cr\lower2.5pt\hbox{$\sim$}}}}}
\def\gsim{\mathrel{\mathstrut\smash{\ooalign{\raise2.5pt\hbox{$>$}\cr\lower2.5pt\hbox{$\sim$}}}}}

\def\overleftrightarrow#1{\vbox{\ialign{##\crcr
     $\leftrightarrow$\crcr\noalign{\kern-0pt\nointerlineskip}
     $\hfil\displaystyle{#1}\hfil$\crcr}}}
     
     \def\overleftarrow#1{\vbox{\ialign{##\crcr
     $\leftarrow$\crcr\noalign{\kern-0pt\nointerlineskip}
     $\hfil\displaystyle{#1}\hfil$\crcr}}}

\usepackage[yyyymmdd,hhmmss]{datetime}
\newif{\ifeq}           
\eqtrue                 
\newcounter{lecturecounter}